\documentclass[aps,prb,twocolumn,amsmath,amssymb,nofootinbib,
  superscriptaddress]{revtex4-2}

\usepackage{graphicx}
\usepackage{bm}
\usepackage{amsmath}
\usepackage{amssymb}
\usepackage{physics}
\usepackage[colorlinks=true,linkcolor=blue,urlcolor=blue,citecolor=blue]{hyperref}
\usepackage[usenames,svgnames]{xcolor}
\usepackage{makecell}
\usepackage[normalem]{ulem}

\newcommand{\Hop}{\hat{H}}
\newcommand{\Himp}{\hat{H}_{\rm imp}}
\newcommand{\Hint}{\hat{H}_{\rm int}}

\newcommand{\Hbath}{\hat{H}_{\rm bath}}
\newcommand{\Hhyb}{\hat{H}_{\rm hyb}}

\newcommand{\aop}{\hat{a}}
\newcommand{\adop}{\hat{a}^{\dagger}}

\newcommand{\cop}{\hat{c}}
\newcommand{\cdop}{\hat{c}^{\dagger}}
\newcommand{\hc}{{\rm H.c.}}
\newcommand{\rhoop}{\hat{\rho}}
\newcommand{\Zimp}{Z_{{\rm imp}}}
\newcommand{\mea}{\mathcal{D}}
\newcommand{\gK}{\mathcal{K}}
\newcommand{\gI}{\mathcal{I}}
\newcommand{\bolda}{\bm{a}}
\newcommand{\boldabar}{\bar{\bm{a}}}
\newcommand{\abar}{\bar{a}}
\newcommand{\im}{{\rm i}}
\newcommand{\contour}{\mathcal{C}}
\newcommand{\gA}{\mathcal{A}}

\newcommand{\gV}{\mathcal{V}}
\newcommand{\boldeta}{\bm{\eta}}
\newcommand{\boldetabar}{\bar{\bm{\eta}}}
\newcommand{\etabar}{\bar{\eta}}
\newcommand{\parity}{\mathcal{P}}
\newcommand{\current}{\mathcal{J}}

\newcommand{\hnu}{Key Laboratory of Low-Dimensional Quantum Structures and Quantum Control of Ministry of Education, Department of Physics and Synergetic Innovation Center for Quantum Effects and Applications, Hunan Normal University, Changsha 410081, China
}

\begin{document}

\title{Grassmann Time-Evolving Matrix Product Operators for Quantum Impurity Problems}

\author{Ruofan Chen}
\affiliation{College of Physics and Electronic Engineering, and Center for Computational Sciences, Sichuan Normal University, Chengdu 610068, China}

\author{Xiansong Xu}
\affiliation{Science, Mathematics and Technology Cluster, Singapore University of Technology and Design, 8 Somapah Road, Singapore 487372}
\affiliation{College of Physics and Electronic Engineering, and Center for Computational Sciences, Sichuan Normal University, Chengdu 610068, China}

\author{Chu Guo}
\email{guochu604b@gmail.com}

\affiliation{\hnu}



\begin{abstract}


The time-evolving matrix product operators (TEMPO) method,  which makes full use of the Feynman-Vernon influence functional, is the state-of-the-art tensor network method for bosonic impurity problems. However, for fermionic impurity problems the Grassmann path integral prohibits application of this method.
We develop Grassmann time-evolving matrix product operators, a full fermionic analog of TEMPO, that can directly manipulates Grassmann path integrals with similar numerical cost as the bosonic counterpart.
We further propose a zipup algorithm to compute expectation values on the fly without explicitly building a single large augmented density tensor,
which boosts our efficiency on top of the vanilla TEMPO. 
Our method has a favorable complexity scaling over existing tensor network methods, and we demonstrate its performance on the non-equilibrium dynamics of the single impurity Anderson models. Our method solves the long standing problem of turning Grassmann path integrals into efficient numerical algorithms, which could significantly change the application landscape of tensor network based impurity solvers, and could also be applied for broader problems in open quantum physics and condensed matter physics.
\end{abstract}

\maketitle


\section{Introduction}

Quantum impurity problems (QIPs), where the impurity is coupled to fermionic baths, are prototypical models to study open quantum dynamics and quantum transport. Besides the fundamental interest in themselves, the Anderson impurity problem~\cite{anderson1961-localized}, in particular, is also a crucial subtask to solve in the dynamical mean field theory (DMFT) for high-dimensional fermionic systems~\cite{GeorgesRozenberg1996}. As such developing efficient numerical methods for QIPs has been a long-lasting pursuit in condensed matter physics.

The past decades have witnessed an outburst of new efficient methods for QIPs. Among them the most outstanding one is perhaps the class of continuous time quantum Monte Carlo methods~\cite{GullWerner2011}. These methods make full use of the analytical properties of the Green's function during the calculation, and are free of the bath discretization and time discretization errors, as such they could often be highly accurate and efficient~\cite{RubtsovLichtenstein2004,RubtsovLichtenstein2005,WernerMillis2006,GullTroyer2008}. They have been widely applied to study QIPs with multiple orbitals~\cite{WernerMillis2006b,Haule2007,WernerMillis2007,WernerMillis2008,ChanMillis2009}, and in non-equilibrium scenarios~\cite{AokiWerner2014}. However, they suffer from sampling noises, the sign problem~\cite{TroyerWiese2005} and the ill-posed analytical continuation when working in the imaginary frequency axis~\cite{WolfSchollwock2015,FeiGull2021}.

Tensor network based methods are another important class of methods which could in principle solve QIPs in a numerically exact way. 
Earlier developments of these methods have adapted the strategy to explicitly discretize the bath and consider the impurity and bath dynamics as a whole~\cite{BullaPruschke2008,WolfSchollwock2014b,WolfSchollwock2014,GanahlEvertz2014,GanahlVerstraete2015,KohnSantoro2021,KohnSantoro2022}.
General advantages of tensor network based methods include that they are free of sampling noises and the sign problem, and that they can directly work in the real frequency axis,
thus avoiding the usage of the numerically ill-posed analytical continuation. 
However, the need to explicitly treat the bath could often result in spatial discretization error and very rapid growth of the computational cost, particularly when in presence of multiple baths~\cite{MitchellBulla2014,StadlerWeichselbaum2015,HorvatMravlje2016,KuglerGeorges2020}.

A key strategy to improve the performance of tensor network based impurity solvers is to explicitly make use of the Feynman-Vernon influence functional~\cite{FeynmanVernon1963} (IF) in the tensor network. This idea has been pioneered for bosonic impurity problems, referred to as the time-evolving matrix product operators (TEMPO) method~\cite{StrathearnLovett2018}. TEMPO represents the IF as a matrix product state (MPS) and constructs it using standard 
MPO (matrix product operator)-MPS arithmetics~\cite{Schollwock2011}. It has been proven to be elegant and efficient, and become state-of-the-art~\cite{joergensen2019-exploiting,popovic2021-quantum,fux2021-efficient,gribben2021-using,otterpohl2022-hidden,gribben2022-exact}. However, TEMPO can not be directly applied to fermionic case due to the presence of Grassmann variables (GVs).

In this work we propose the Grassmann time-evolving matrix product operators (GTEMPO) for QIPs, which is an extension of TEMPO to fermions. 
The key challenge of dealing with GVs numerically is addressed
by representing the Grassmann tensors (GTs) as Grassmann MPSs (GMPSs) that respect the anti-commutation relation, which allows us to transplant all the major techniques of TEMPO to the fermionic case with the same numerical effort as the bosonic counterparts. 
We further propose a zipup algorithm to compute expectation values on the fly without explicitly building the augmented density tensor as a single GMPS, but based on several GMPSs for the bare impurity dynamics and the IF respectively. This algorithm greatly boosts the efficiency of our method compared to the original TEMPO, and allows us to exactly treat the bare impurity dynamics with negligible cost.
Our method significantly differs from the recently proposed tensor network IF method~\cite{ThoennissAbanin2023a,ThoennissAbanin2023b}.
The latter approach reverts the Grassmann expression of the fermionic path integral (PI) back into the evaluation of operator expectation values and treats the single-bath and two-bath impurity problems separately, while we use GMPS to directly represent the integrand of the PI and our implementation can be readily used for any impurity problems. It also uses the Fishman-White algorithm~\cite{fishman2015-compression} to build the IF as an MPS (MPS-IF), which introduces an additional hyperparameter to monitor convergence. In comparison our method uses GMPS multiplications in which the only hyperparameter is the bond truncation tolerance of MPS (throughout this work we use the same strategy as in Ref.~\cite{StrathearnLovett2018} for MPS bond truncation, i.e., we truncate all the singular values with relative weights lower than the tolerance). Besides, it constructs one MPS-IF per spin species per bath, while our method only needs to construct one MPS-IF per spin species for all baths. Since the computational cost scales exponentially with the number of MPS-IFs, it could be difficult to apply the tensor network IF method to study more complex impurity models.
The performance of our method is demonstrated on the the non-equilibrium dynamics of the single impurity Anderson models (SIAMs), where we show that we can achieve higher accuracy and efficiency than the tensor network IF method. 
\section{Model description}

The Hamiltonian for general QIPs can be written as
$\Hop = \Himp + \Hbath + \Hhyb$, with $\Himp$, $\Hbath$ and $\Hhyb$
the impurity, bath and hybridization Hamiltonians respectively. 
In the following, we will use the real-time evolution of the single-bath SIAM from a product impurity-bath initial state as a concrete model to illustrate the basic principles of our method,
for which $\Himp = (\epsilon_d-\frac{1}{2}U)\sum_{\sigma}\adop_{\sigma}\aop_{\sigma} + U \adop_{\uparrow}\aop_{\uparrow}\adop_{\downarrow}\aop_{\downarrow}$, $\Hbath = \sum_{k, \sigma}\epsilon_k \cdop_{k, \sigma}\cop_{k, \sigma}$, and $\Hhyb =\sum_{k, \sigma}\left(V_k \adop_{\sigma}\cop_{k, \sigma} + \hc \right)$,
with $\epsilon_d$ the on site energy of the impurity, $U$ the Coulomb interaction, $\epsilon_k$ the band energy and $V_k$ the hybridization strength. The initial state is $\rhoop(0) = \rhoop_{{\rm imp}}(0) \otimes \rhoop_{{\rm bath}}^{{\rm th}}$, with $\rhoop_{{\rm imp}}(0)$ the impurity initial state and $\rhoop_{{\rm bath}}^{{\rm th}}$ the bath equilibrium state.
The PI of the impurity partition function $\Zimp(t)=\Tr\rhoop(t)/\Tr \rhoop_{\mathrm{bath}}^{\mathrm{th}}$ at time $t$ can be written in terms of Grassmann trajectories as~\cite{kamenev2009-keldysh,negele1998-quantum}
\begin{align}\label{eq:PI}
\Zimp(t) = \int \mathcal{D}[\boldabar,\bolda] \gK\left[\boldabar, \bolda \right]\prod_{\sigma}\gI_{\sigma}\left[\boldabar_{\sigma}, \bolda_{\sigma}\right],
\end{align}
where  $\boldabar_{\sigma} = \{\abar_{\sigma}(\tau)\}$, $\bolda_{\sigma} = \{a_{\sigma}(\tau)\},\boldabar=\{\boldabar_{\uparrow},\boldabar_{\downarrow}\},\bolda=\{\bolda_{\uparrow},\bolda_{\downarrow}\}$ for briefness, and the measure $\mathcal{D}[\boldabar,\bolda]=\prod_{\sigma,\tau}\dd\abar_{\sigma}(\tau)\dd a_{\sigma}(\tau)
  e^{-\abar_{\sigma}(\tau)a_{\sigma}(\tau)}$ ($\abar$ denotes the conjugate GV of $a$).
$\gK[\boldabar, \bolda]$ is the bare impurity propagator, and $\gI_{\sigma}[\boldabar_{\sigma}, \bolda_{\sigma}]$ denotes the IF for a single spin species:
\begin{align}\label{eq:I}
\gI_{\sigma}[\boldabar_{\sigma}, \bolda_{\sigma}] = e^{-\int_{\contour} \dd\tau \int_{\contour}\dd\tau' \abar_{\sigma}(\tau) \Delta(\tau, \tau') a_{\sigma}(\tau') },
\end{align}
where $\mathcal{C}$ is the Keldysh contour. The hybridization function $\Delta(\tau, \tau')$ completely characterizes the effect of the bath, which is determined by the bath spectrum density $J(\omega)$ and free bath Green's function $D_{\omega}(\tau,\tau')$ that depends on the inverse temperature $\beta$ and the chemical potential $\mu$.

\subsection{Discretization of the path integral}

For numerical calculation the integration in the PI should be discretized. Breaking $[0, t]$ into $M$ discrete points with equal-distant interval $\delta t=t/(M-1)$, and denoting $\Hint = \Hhyb + \Hbath$, a first-order discretized expression for Eq.(\ref{eq:PI}) can be straightforwardly derived by inserting the Grassmann identity operator~\cite{kamenev2009-keldysh,negele1998-quantum} at each discretized time step of the following first-order trotter decomposition of the evolutionary operator $e^{-\im \Hop t}$:
\begin{align}
e^{-\im\Hop t} &= e^{-\im\Hop \delta t} \cdots e^{-\im\Hop \delta t} \nonumber \\ 
&\approx e^{-\im\Hint \delta t} e^{-\im\Himp \delta t} \cdots e^{-\im\Hint \delta t} e^{-\im\Himp \delta t}, 
\end{align}
where in the second line each term $e^{-\Hop \delta t}$ is approximated by its first-order expression $e^{-\im\Hint \delta t} e^{-\im\Himp \delta t}$. One can also use a symmetric trotter decomposition as $e^{-\im \Hop \delta t} \approx e^{-\im\Hint \delta t/2} e^{-\im\Himp \delta t} e^{-\im\Hint \delta t/2} $~\cite{trotter1959-product,suzuki1976-generalized}, which leads to a second-order decomposition of $e^{-\im\Hop t} $:
\begin{align}
e^{-\im\Hop t} \approx  &e^{-\im\Hint \delta t/2} e^{-\im\Himp \delta t} e^{-\im\Hint \delta t} \times \cdots \nonumber \\ 
&\times e^{-\im\Himp \delta t} e^{-\im\Hint \delta t/2}.
\end{align}
For numerical calculation with finite $\delta t$, the first-order and second-order decompositions can make a difference, which will be referred to as the first-order GTEMPO and second-order GTEMPO respectively.

Discretizing $\contour$ results in $4M$ discrete GVs per spin species (this number is $8M$ in Ref.~\cite{ThoennissAbanin2023b}):
half for the forward branch ($+$) and half for the backward branch ($-$), denoted as $\bolda^{\pm}_{\sigma} = \{a_{\sigma, M}^{\pm}, \cdots, a_{\sigma, 1}^{\pm}\} $ and $\boldabar^{\pm}_{\sigma} = \{\abar_{\sigma, M}^{\pm}, \cdots, \abar_{\sigma, 1}^{\pm}\}$ respectively.
Under these notations, the double integral in Eq.~(\ref{eq:I}) can be discretized via the quasi-adiabatic propagator path integral (QuaPI) method~\cite{makarov1994-path,makri1995-numerical,dattani2012-analytic} as $\int_{\contour} \dd\tau \int_{\contour}\dd\tau' \abar_{\sigma}(\tau) \Delta(\tau, \tau') a_{\sigma}(\tau') \approx \sum_{\zeta,\zeta'=\pm}\sum_{jk}\abar^{\zeta}_{\sigma,j}\Delta_{jk}^{\zeta \zeta'}a^{\zeta'}_{\sigma,k}$ (See Appendix.~\ref{app:quapi} for details of QuaPI and Appendix.~\ref{app:K} for discretization of $\gK$). The discretized integrand of Eq.(\ref{eq:PI}) is referred to as the augmented density tensor (ADT)~\cite{StrathearnLovett2018}, denoted as
\begin{align}\label{eq:adt}
\gA[\boldabar, \bolda] = \gK[\boldabar, \bolda] \prod_{\sigma} \gI_{\sigma}[\boldabar, \bolda].
\end{align}


\section{Method description}\label{sec:method}
Before we describe our method in detail, we briefly review the central idea of TEMPO first. TEMPO represents the ADT as an MPS, which includes both the effects of the impurity dynamics $\gK$ and the IF $\gI$.  In particular, $\gI$ is built as an MPS by multiplying $O(M)$ matrix product operators (MPOs), each representing a ``partial'' IF (which will be defined later), onto the vacuum state. The observables are calculated by inserting operators into appropriate positions in the PI and integrating the resulting expression. In short, TEMPO directly translates the textbook PI formalism into efficient MPS calculations. Crucially, since only the variables representing the impurity state at different time steps appear in the PI, the implementation of TEMPO is universal for different types of baths and its computational cost is only directly dependent on the number of impurities.
More concretely, if there are $n$ baths coupled to the same impurity in the same form as $\Hhyb$, with hybridization functions denoted as $\Delta^{\nu}(\tau, \tau')$ for $1\leq \nu\leq n$, then at the level of the PI (thus for the TEMPO method), their effect is completely equivalent to \textit{one effective bath} with the hybridization function $\Delta(\tau, \tau') = \sum_{\nu=1}^n \Delta^{\nu}(\tau, \tau')$. This is a major difference of TEMPO compared to those tensor network IF methods in Refs.~\cite{ThoennissAbanin2023a,ThoennissAbanin2023b,NgReichman2023} in terms of computational cost.

The major difficulty of generalizing TEMPO to fermionic impurity problems is that fermionic PI contains GVs (which are normal numbers in the bosonic case). For this we first introduce the concept the Grassmann Tensors. 
For an algebra of GVs of $n$ components
$\xi_1,\ldots,\xi_n$, we can define the GT as
\begin{align}
  \gA = \sum_{\bm{i}} A^{i_n, \ldots, i_1} \xi^{i_n}_n \cdots \xi^{i_1}_1,\quad i_k=\{0,1\},
\end{align}
with $\bm{i} = \{i_n, \cdots, i_1\} $. 
The coefficient tensor $A^{i_n, \cdots, i_1}$ is a normal rank-$n$ array of $2^n$ scalars.
The discretized $\gK$ and $\gI_{\sigma}$ are essentially GTs
spanned by $\{\boldabar,\bolda\}$. 
In principle, any GT can be constructed from smaller
pieces of GTs by their multiplications (Grassmann tensor multiplication is a generalization of the standard tensor multiplication to GTs, which can be understood as the outer multiplication of two GTs followed by contraction of common tensor indices). Unfortunately, GT multiplication does not simply boil down to multiplication of their coefficient tensors, since one has to first bring the same GVs of the two GTs into nearby positions. 
This operation requires a series of swap operations to take into account the sign changes (One can refer to a recent work for an implementation of GT multiplications which explicitly takes into account the sign changes~\cite{Yosprakob2023}).
The situation is even worse if one directly represents the coefficient tensor as an MPS, 
since long range swap operations on an MPS could in the worst case induce an exponential increase of its bond dimension.

\begin{figure}
  \includegraphics[width=\columnwidth]{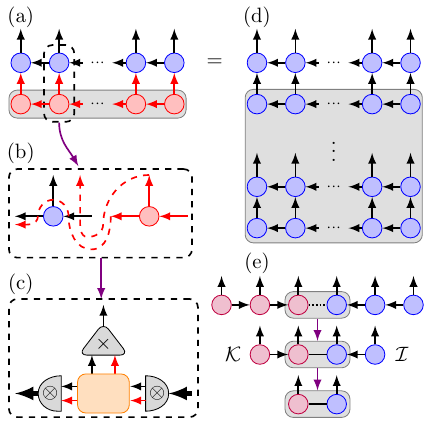} 
  \caption{(a) Multiplication of two GMPSs in a site-by-site manor as the bosonic case. Multiplication of two Grassmann site tensors is done in two steps: (b) Tensor product of two site tensors to obtain a rank-$6$ tensor, followed by a permutation of the tensor indices (the local GVs are reordered accordingly) as indicated by the dashed red lines; and (c) Regrouping the tensor indices to restore a rank-$3$ site tensor, where the auxiliary indices are fused together while the physical indices are multiplied together.
  (d) GTEMPO algorithm to build the MPS-IF using a series of GMPS multiplications corresponding to Eq.~(\ref{eq:partialIF}). 
  (e) The zipup algorithm to compute expectation values exemplified for the case of two GMPSs for $\gK$ and $\gI$, where the multiplication of two GMPSs is performed site by site to integrate the ADT on the fly.
    }
    \label{fig:fig1}
\end{figure}

Our solution is to represent the GTs as GMPSs:
\begin{align}\label{eq:gmps}
\gA = &\sum_{\bm{i}, \bm{\alpha}, \bar{\bm{\alpha}}}A^{i_n}_{\alpha_{n-1}} \cdots A^{i_j}_{\bar{\alpha}_j, \alpha_{j-1}}\cdots A^{i_1}_{\bar{\alpha}_1} \times \nonumber \\
& \int \mea\left[\boldetabar, \boldeta\right]  \xi^{i_n}_n \eta^{\alpha_{n-1}}_{n-1}\cdots \etabar_{j}^{\bar{\alpha}_j} \xi^{i_j}_j \eta_{j-1}^{\alpha_{j-1}}\cdots \etabar_{1}^{\bar{\alpha}_1} \xi^{i_1}_1,
\end{align}
where we have enlarged the Grassmann space by inserting two GVs $\etabar_{k},\eta_k$ between $\xi_k$ and $\xi_{k+1}$.
The auxiliary index $\alpha_j$ should be understood as a tuple $(\alpha_j, m_{\alpha_j})$ with $\alpha_j$ the exponent of $\eta_j$ and $m_{\alpha_j}$ the inner degeneracy index corresponding to $\alpha_j$.
Since $\gK$ and $\gI_{\sigma}$ always contain an even number of GVs, 
we can require each site tensor $A^{i_k}_{\bar{\alpha}_k, \alpha_{k-1}}$ in Eq.~(\ref{eq:gmps}) to satisfy the \textit{even parity condition} (with $\alpha_0 = \bar{\alpha}_n = 0$):
\begin{align}\label{eq:evenparity}
\mod(\bar{\alpha}_{k} + i_k + \alpha_{k-1}, 2) = 0.
\end{align}
With Eqs.~(\ref{eq:gmps},\ref{eq:evenparity}) the GMPS can be naturally implemented as a $Z_2$-symmetric MPS (See Ref.~\cite{SinghVidal2011} for example for MPSs with Abelian symmetries).
We note that the fermionic MPSs~\cite{FidkowskiKitaev2011,bultinck2017-fermionic}, as well as the higher dimensional Grassmann tensor networks~\cite{gu2013-efficient,yoshimura2018-calculation,akiyama2021-more}, share some similarity with our GMPS. 
Eq.~\eqref{eq:evenparity} ensures that the local GVs $\etabar_k^{\bar{\alpha}_k}\xi^{i_k}\eta_{k-1}^{\alpha_{k-1}}$ have even parity, thus they can be freely moved around as a whole without any sign changes. This property allows to multiply two GMPSs 
in a site-by-site fashion \textit{without swap operations}, as depicted in Fig.~\ref{fig:fig1}(a). The arrows on the legs of the GMPS represent the flow of good quantum numbers (parity in our case) as a general feature of symmetry protected MPS.
Multiplying two site tensors $A^{i_k}_{\bar{\alpha}_k, \alpha_{k-1}} $ and $B^{i_k'}_{\bar{\alpha}_k', \alpha_{k-1}'}$ can be done in two steps. First, performing standard bosonic tensor product together with a permutation of tensor indices as shown in Fig.~\ref{fig:fig1}(b), resulting in a rank-$6$ tensor $(-1)^{i_k'\alpha_{k-1} } A^{i_k}_{\bar{\alpha}_k, \alpha_{k-1}} \otimes B^{i_k'}_{\bar{\alpha}_k', \alpha_{k-1}'}$, where the sign is due to the reordering of the local GVs from $\etabar_k\xi_k\eta_{k-1}\etabar_k'\xi_k'\eta_{k-1}'$ to $\etabar_k'\etabar_k \xi_k \xi_k' \eta_{k-1}\eta_{k-1}'$. Second, regrouping the physical and auxiliary indices to restore the rank-$3$ site tensor as shown in Fig.~\ref{fig:fig1}(c), where the auxiliary indices are grouped by bosonic fusion operation of $Z_2$-symmetric tensor indices~\cite{SinghVidal2011}, while Grassmann multiplication is performed for the physical indices since $\xi_k = \xi_k'$. 
The multiplication ends with a bond truncation of the resulting GMPS, which can be done in the same way as the bosonic case (details can be found in Appendix.~\ref{app:sitemult}).

With GMPS it is straightforward to transplant the techniques of TEMPO to deal with fermionic PIs.
For this we first rewrite the IF for a fixed $\zeta$ and $\sigma$ as 
\begin{align}\label{eq:partialIF}
e^{-\sum_{j,k,\zeta'}\abar_{\sigma,j}^{\zeta} \Delta^{\zeta\zeta'}_{jk}a_{\sigma,k}^{\zeta'} }= \prod_{j} e^{-\abar_{\sigma, j}^{\zeta} \sum_{k,\zeta'} \Delta^{\zeta\zeta'}_{jk} a_{\sigma, k}^{\zeta'}},
\end{align}
The operand on the rhs of Eq.~\eqref{eq:partialIF} is a partial IF, whose exponent contains quadratic terms of GVs that share the same $\abar_{\sigma, j}^{\zeta}$. Similar to the bosonic case, this partial IF can be compactly written as a GMPS with a small fixed bond dimension (See Appendix.~\ref{app:I} for detailed construction). As a result, one only needs $O(M)$ GMPS multiplications to build the MPS-IF as shown in Fig.~\ref{fig:fig1}(d).
In comparison, $O(M^2)$ operations are required if Eq.~\eqref{eq:partialIF} is broken into products of two-body terms like $e^{\lambda\xi_i\xi_j}$.

$\gK$ can be built as a GMPS 
by multiplying $O(M)$ two-body and four-body ($e^{\lambda\xi_i\xi_j\xi_k\xi_l}$) terms, which can be done exactly without bond truncation and with negligible cost compared to $\gI_{\sigma}$. After that, the ADT can be obtained using Eq.(\ref{eq:adt})
as done in TEMPO. 
With ADT one can easily compute expectation values, for example the discretized Green's function can be computed as
\begin{align}\label{eq:gf}
\langle\aop_k \adop_j \rangle = \langle a_k \abar_j \rangle := Z_{{\rm imp}}^{-1}\int \mathcal{D}[\boldabar,\bolda] a_k \bar{a}_j \mathcal{A}[\boldabar,\bolda].
\end{align}
However the bond dimension of the ADT will be much larger than that of individual $\gI_{\sigma}$ or $\gK$ in general. We propose a zipup algorithm to compute expectation values on the fly based on multiple GMPSs instead of only one GMPS for the ADT, as exemplified in Fig.~\ref{fig:fig1}(e) for two GMPSs $\gK$ and $\gI$. For SIAMs with two spin species (thus we have three GMPSs, $\gK$, $\gI_{\uparrow}$ and $\gI_{\downarrow}$ in total), this algorithm drastically decreases the computational cost of time evolution from $O(M\chi_{\mathcal{A}}^3) $ to $O(M\chi^3)$ (neglecting the cost of building $\gK$), and the cost of computing expectation values from $O(M\chi_{\gA}^3)$ to $O(M\chi^3\chi_{\gK})$, with $\chi$, $\chi_{\gK}$ and $\chi_{\mathcal{A}}$ the bond dimensions of $\gI_{\sigma}$, $\gK$ and the ADT respectively, and $\chi_{\gA} = \chi^2\chi_{\gK}$ in the worst case. The memory cost of our method scales as $O(\chi^2\chi_{\gK})$ only, which is the size of the largest intermediate tensor during computing observables. We note that our method has a much more favorable scaling compared to Refs.~\cite{ThoennissAbanin2023b} (whose memory cost scales as $O(\chi^4\chi_{\gK})$ for SIAM with two baths) since we need only one MPS-IF per spin species, irrespective of the number of baths. In the meantime, we stress that for higher-orbital impurity models, the computational cost of our method will still scale exponentially as the number of orbitals, since the number of MPS-IFs scales as two times the number of orbitals (the factor of two is for the two spins).

The ordering of GVs could also significantly affects the bond dimension of $\mathcal{K}$ and $\mathcal{I}_{\sigma}$. We have adapted an ordering that the GVs at different time steps are aligned in descending order, while those within the same time step $j$ are grouped together~\cite{BanulsCirac2009,KohnSantoro2022} and aligned as $a_{\uparrow, j}^+a_{\uparrow, j}^- a_{\downarrow, j}^+a_{\downarrow, j}^- \abar_{\downarrow, j}^- \abar_{\downarrow, j}^+ \abar_{\uparrow, j}^- \abar_{\uparrow, j}^+ $, which gives us good performance in general.


\section{Numerical results}

To demonstrate our method we apply it to numerically study the SIAMs in the non-interacting limit, with a single bath and with two baths respectively. In the non-interacting case, the accuracy of our method is benchmarked against the analytical solutions and the errors related to the two hyperparameters of our method are analyzed. In the rest two cases we benchmark our results to existing calculations using tensor network IF methods and quantum Monte Carlo methods.
For all our numerical calculations we will focus on the bath spectrum density 
\begin{align}\label{eq:spectrum}
J(\omega) = \frac{\Gamma}{2\pi} D\sqrt{1 - (\omega/D)^2}
\end{align}
with  $\Gamma = 0.1$. The impurity initial state is chosen as $\rhoop_{{\rm imp}}(0) = \ketbra{0}{0}$.

\subsection{The non-interacting case}

\begin{figure}
  \includegraphics[width=\columnwidth]{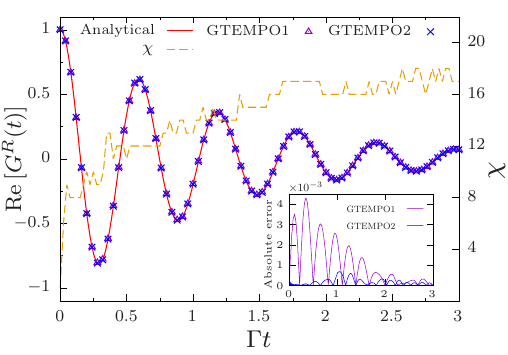} 
  \caption{The non-equilibrium retarded Green's function for the non-interacting SIAM (left axis) computed using first-order GTEMPO (GTEMPO1) and second-order GTEMPO (GTEMPO2) methods, compared to the analytical solutions. The yellow dash line (right axis) shows the growth of the bond dimension of the MPS-IF for TEMPO2 during the time evolution. We have used $\Gamma\delta t=0.005$, $D=2$ and $\varsigma = 10^{-7}$ for both methods. The inset shows the absolute errors of the GTEMPO results with the analytical solutions.
    }
    \label{fig:fig2}
\end{figure}

We first apply GTEMPO to calculate the analytically solvable non-equilibrium retarded Green's function $G^R(t)=\Theta(t)\expval*{\aop_{\sigma}(t)\adop_{\sigma}+\adop_{\sigma}\aop_{\sigma}(t)}$ of the single-bath SIAM at $U=0$, where only one spin species needs to be considered.
We expect the accuracy of our method for $U \neq0$ is similar to that of $U=0$ since $\gK$ is treated exactly (except that for $U \neq0$ one may need a smaller $\delta t$).
In Fig.~\ref{fig:fig2}, we benchmarked both the first and second-order GTEMPO methods with the analytical solutions till $\Gamma t=3$, where we have chosen $D=2$ and the bond truncation tolerance as $\varsigma = 10^{-7}$. The inset shows the absolute errors of these methods against the analytical solution. We can see that both methods give very accurate results. The first-order method is less accurate in the beginning than the second-order method, but eventually the accuracy of both methods becomes similar (likely limited by the bond truncation tolerance). We can also see that the bond dimension $\chi$ increases to around $16$ (at most $18$) at $\Gamma t \approx 1.5$ and then almost stops growing, which means that we can easily simulate long time dynamics with moderate numerical effort.

\begin{figure}
  \includegraphics[width=\columnwidth]{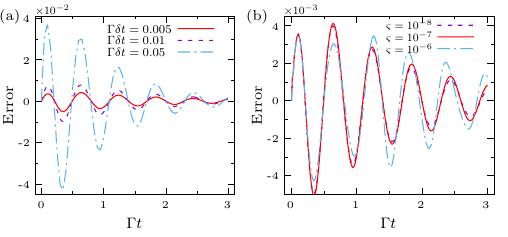} 
  \caption{The error of first-order GTEMPO calculations against the analytical solutions, for the real part of the non-equilibrium retarded Green's function of the non-interacting SIAM. In panel (a) we have fixed $\varsigma=10^{-7}$ and considered $\Gamma \delta t=0.005, 0.01, 0.05$, while in panel (b) we have fixed $\Gamma \delta t =0.005$ and considered $\varsigma=10^{-6}, 10^{-7}, 10^{-8}$.
    }
    \label{fig:fig3}
\end{figure}

Before we move on to more sophisticated impurity models, we analyze the errors of our GTEMPO calculations with respect to the two hyperparameters, $\delta t$ and $\varsigma$, for the non-interacting case in Fig.~\ref{fig:fig3}. From Fig.~\ref{fig:fig3}(a), we can clearly see that our GTEMPO results becomes more accurate with smaller $\delta t$. From Fig.~\ref{fig:fig3}(b), we can see that the error for the non-interacting case remains more or less the same when decreasing $\varsigma$ from $10^{-6}$ to $10^{-8}$ for fixed $\Gamma \delta t=0.005$, thus the dominate source of error in this simulation is really the time discretization error. We can also see that the error in the GTEMPO results does not cumulate with time, in comparison with the wave function based time evolution methods~\cite{PaeckelHubig2019}.

\begin{figure}
  \includegraphics[width=\columnwidth]{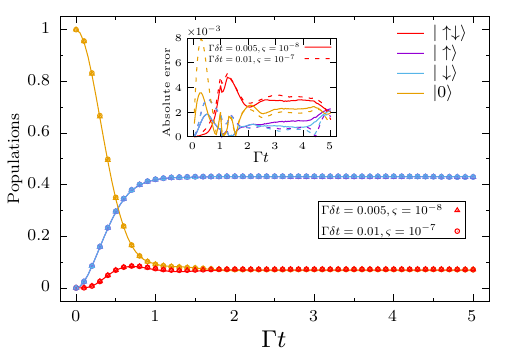} 
  \caption{Time evolution of the populations of the impurity for the single-bath SIAM. The solid lines are results taken from Ref.~\cite{NgReichman2023}. Two sets of first-order GTEMPO calculations, with $\Gamma\delta t=0.01$, $\varsigma=10^{-7}$ (marked with circle) and $\Gamma\delta t=0.005$, $\varsigma=10^{-8}$ (marked with triangle), are shown. The inset shows the errors of these two sets of GTEMPO calculations against results from Ref.~\cite{NgReichman2023}. For this simulation we have used $D=1$ and $\Gamma\beta=2$.}
    \label{fig:fig4}
\end{figure}

\subsection{Equilibration dynamics}\label{sec:singlebath}

Next we study the equilibration dynamics of the single-bath SIAM with $D=1$ and $\Gamma\beta=2$. In Fig.~\ref{fig:fig4}, we show the time evolution of the populations of the impurity in the four states $\vert \uparrow \downarrow\rangle$ (double occupancy), $\vert \uparrow\rangle$ (spin up), $\vert \downarrow\rangle$ (spin down) and $\vert 0\rangle$ (no electron), and compare our first-order GTEMPO results to existing calculations from Ref.~\cite{NgReichman2023}. We have performed two sets of GTEMPO calculations with $\Gamma\delta t=0.01$, $\varsigma=10^{-7}$ and $\Gamma\delta t=0.005$, $\varsigma=10^{-8}$ respectively (it has been shown that for TEMPO methods a general rule of thumb is that smaller $\varsigma$ should be used with smaller $\delta t$~\cite{otterpohl2022-hidden}).
We can see that the errors of both our calculations compared to Ref.~\cite{NgReichman2023} are of the order $10^{-3}$, which show that our GTEMPO calculations are very accurate and have well converged. We observe that the largest bond dimensions of the MPS-IFs in these two sets of calculations are $64$ and $67$ respectively.

\subsection{Non-equilibrium steady state}

Finally we apply our method to study the non-equilibrium transport of the SIAM with two baths~\cite{BertrandWaintal2019,ThoennissAbanin2023b}. The only change that needs to be made for our method, compared to the case of one bath, is that the hybridization function becomes $\Delta(\tau, \tau') = \sum_{\nu=1,2} \Delta^{\nu}(\tau, \tau')$ with $\Delta^{\nu}(\tau, \tau')$ the hybridization function for the $\nu$-th bath, therefore the presence of two baths does not affect the computational cost for our method in principle.
The particle current with spin $\sigma$ flows out of the $\nu$-th bath, defined as
\begin{align}\label{eq:currenta}
\current^{\nu}_{\sigma}(t) = -\dd\expval*{\hat{N}_{\sigma}^{\nu}(t)} / \dd t,
\end{align}
with $\hat{N}_{\sigma}^{\nu}$ is the corresponding particle number operator for the $\nu$-th bath, can be computed via an integral on the Keldysh contour as 
\begin{align}\label{eq:currentb}
  \current^{\nu}_{\sigma}(t) = -2\mathrm{Re}\int_{\mathcal{C}} \dd\tau\Delta^{\nu}(t^+,\tau)\expval*{\abar_{\sigma}(t^+)a_{\sigma}(\tau)},
\end{align}
where $t^+$ means time $t$ on the forward branch. Eq.(\ref{eq:currentb}) can then be discretized into the summation of quadratic terms using a method similar to QuaPI, and each quadratic term can be evaluated using Eq.(\ref{eq:gf}). The derivation of the particle current expression as well its discretization can be found in Appendix.~\ref{app:current}.

\begin{figure}
  \includegraphics[width=\columnwidth]{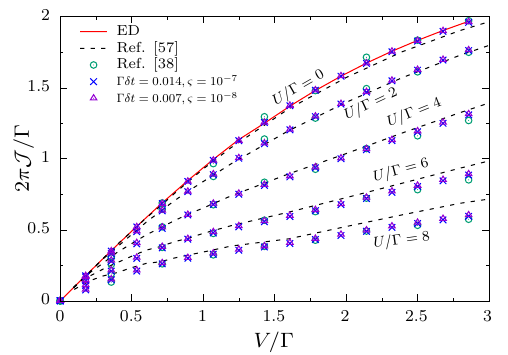} 
  \caption{Current vs voltage for the two-bath SIAM described in the main text for different $U$, calculated by ED (red solid line for $U=0$), first-order GTEMPO with $\Gamma\delta t=0.014$ and $\varsigma=10^{-7}$ (blue x) and with  $\Gamma\delta t=0.007$ and $\varsigma=10^{-8}$ (purple triangle), state-of-the-art Monte Carlo~\cite{BertrandWaintal2019} (black dashed lines) and tensor network IF~\cite{ThoennissAbanin2023b} (green circles). We have used a Trotter step $\delta t=0.007/\Gamma$, and the symmetrized particle current $\current = (\current_{\uparrow}^1 - \current_{\uparrow}^2) / 2 = (\current_{\downarrow}^1 - \current_{\downarrow}^2) / 2 $ is plotted.
    }
    \label{fig:fig5}
\end{figure}

We set the two baths at zero temperature with chemical potentials $\mu_1 = -\mu_2 = V/2$, the total evolution time as $\Gamma t=4.2$ (for fair comparison we have used the same parameter settings as in Ref.~\cite{ThoennissAbanin2023b}).
The steady state current-voltage relation is shown in Fig.~\ref{fig:fig5}, where we have benchmarked the accuracy of our first-order GTEMPO with exact diagonalization (ED), state-of-the-art Monte Carlo calculations~\cite{BertrandWaintal2019}, and the tensor network IF calculations~\cite{ThoennissAbanin2023b}. For ED we have discretized each bath into $4000$ equal-distant frequencies and evolved the system till time $t$ (we have verified that our ED results have well converged against bath discretization). 
Crucially, for $U=0$ our results best agree with ED, which demonstrates that our method is more accurate than existing methods in this case. 
For the $U\neq 0$ cases, it is not possible to draw a decisive conclusion of which method is more accurate since there does not exist exact solutions or widely accepted and very accurate numerical solutions. Nevertheless, we expect that our GTEMPO results are more accurate than the existing calculations based on tensor network IF method for at least four reasons: (1) $\gK$ is treated exactly in our method while the same $\gI_{\sigma}$ is used for all different $U$s (including $U=0$) in our calculations, therefore it is quite likely the accuracy of our method for $U=0$ carries on to the $U\neq 0$ cases; (2) We have shown two sets of calculations with $\Gamma\delta t=0.014$, $\varsigma=10^{-7}$ and with $\Gamma\delta t=0.007$, $\varsigma=10^{-8}$, and we can see that both results converge well with each other for all $U$s; (3) we use Eq.(\ref{eq:currentb}) to calculate the current while Ref.~\cite{ThoennissAbanin2023b} used Eq.(\ref{eq:currenta}), the two expressions are equivalent at the limit $\dd t\rightarrow 0$ but the finite difference expression in Eq.(\ref{eq:currenta}) is more prone to inaccuracies for numerical calculations with finite $\dd t$; (4) In both sets of our GTEMPO calculations we observe that the bond dimension is close to $160$ for all the different parameters considered, while in Ref.~\cite{ThoennissAbanin2023b} a maximum bond dimension $32$ was used, therefore our obtained MPS-IFs are likely to be more accurate.
To this end we also stress that even though we have used a larger bond dimension, the scaling of the computational cost for this case is only $O(\chi^2)$ (See Sec.~\ref{sec:method}) since in our approach we are effectively dealing with a single-bath SIAM as in Sec.~\ref{sec:singlebath}, in comparison the method in Ref.~\cite{ThoennissAbanin2023b} scales as $O(\chi^4)$ in this case. Therefore our method is still more efficient even though a larger bond dimension is used.

\section{Summary}

We have proposed Grassmann time-evolving matrix product operators to solve quantum impurity problems. Similar to TEMPO, our method fully utilizes the analytical expression of the Feynman-Vernon influence functional for fermionic systems and efficiently constructs the MPS-IF using a series of Grassmann matrix product state multiplications. 
As a result the only remaining errors in our method are the time discretization error and the bond truncation error.
We further proposed a zipup algorithm, which computes expectation values on the fly without explicitly building the augmented density tensor as a GMPS, but directly based on several GMPSs for the bare impurity dynamics and the IFs respectively. This technique makes our method much more efficient than the original TEMPO. 
Since our method only depends on the fermionic PI, it works on both the real and imaginary frequency axis, and can be applied to both the equilibrium and non-equilibrium QIPs.
The power of our method is demonstrated on the single impurity Anderson models with one and two baths, where we show that very accurate results can be obtained with moderate bond dimension for the MPS-IF. 
In future investigations it is promising to 
to apply our method to more complex impurity models, as well as to integrate our method in DMFT and non-equilibrium DMFT.

%
%

\acknowledgements 
We thank Nathan Ng for providing data from their simulations.
This work is supported by National Natural Science Foundation of China under Grant No. 12104328. C. G. is supported by the Open Research Fund from State Key Laboratory of High Performance Computing of China (Grant No. 202201-00).

\appendix

\section{Fermionic quasi-adiabatic propagator path integral}\label{app:quapi}

The hybridization function $\Delta(\tau,\tau')$ is computed as
\begin{align}
  \Delta(\tau,\tau')= \parity_{\tau\tau'}\int \dd\omega J(\omega)D_{\omega}(\tau,\tau'),
\end{align}
where $D_{\omega}(\tau,\tau')$ is the free contour-ordered Green's
function of the bath, defined as
\begin{align}
D_{\omega}(\tau,\tau') = \expval*{T_{\mathcal{C}}\cop_{\omega}(\tau)\cdop_{\omega}(\tau')}_0,
\end{align}
with $\parity_{\tau\tau'}=1$ if $\tau$, $\tau'$ are on same Keldysh branch, and $-1$ otherwise.

For numerical calculations, the IF needs to be discretized. In the bosonic case,  it is
already known that a brute-force discretization
could lead to very inaccurate results. Therefore an improved discretization
scheme, which is called the quasi-adiabatic propagator path integral (QuaPI)
\cite{makarov1994-path,makri1995-numerical,dattani2012-analytic},
should generally be adopted. 
The QuaPI method can be straightforwardly generalized to the fermionic case.
For ease of exposition, here we take the first-order
Trotter-Suzuki splitting
\cite{trotter1959-product,suzuki1976-generalized} as example, which can also be adapted to higher order symmetrized Trotter-Suzuki splitting.

In the normal time axis the Grassmann variables are split into two
branches as $\bar{a}_{\sigma}^{\pm}(t),a_{\sigma}^{\pm}(t)$, and  accordingly the
hybridization function is split into four blocks as
\begin{align}
  \Delta(t',t'') = \mqty[\Delta^{++}(t',t'')&\Delta^{+-}(t',t'')\\
  \Delta^{-+}(t',t'')&\Delta^{--}(t',t'')].
\end{align}
We split the trajectories
$\bar{a}_{\sigma}^{\pm}(t),a_{\sigma}^{\pm}(t)$ into intervals of
equal duration as $\bar{a}^{\pm}_{\sigma,j},a_{\sigma,j}^{\pm}$ in
$(j-\frac{1}{2})\delta t<t<(j+\frac{1}{2})\delta t$, then the
hybridization function is discretized as
\begin{align}
  \Delta_{j,k}^{\zeta\zeta'}=\int_{(j-\frac{1}{2})\delta t}^{(j+\frac{1}{2})\delta t}\dd t'
  \int_{(k-\frac{1}{2})\delta t}^{(k+\frac{1}{2})\delta t}\dd t''\Delta^{\zeta\zeta'}(t',t'')
\end{align}
for $\zeta, \zeta' = \pm$ and $1\leq j, k\leq M$.

\section{Derivations of the path integral formalism for the particle current}\label{app:current}

We first consider the single-bath SIAM. The particle current $\current_{\sigma}(t)$ flows out of the bath for spin species $\sigma$ is defined as the opposite of 
particle number change per unit time in the bath:
\begin{align}\label{eq:curren1}
    \current_{\sigma}(t)=- \frac{\dd\expval*{\hat{N}_\sigma(t)}}{\dd t} 
    =2\Im\left(\sum_kV_k\expval*{\hat{a}_{\sigma}^{\dag}(t)c_{\sigma,k}(t)}\right).
\end{align}
In our GTEMPO method, the bath degrees of freedom are traced out and Eq. \eqref{eq:curren1} can not be directly evaluated. Nevertheless, we can convert Eq. \eqref{eq:curren1} into an equivalent expression of the impurity's Green's functions as follows.

We first define a ``modified'' partition function $Z^{\kappa_{\sigma}}_{\mathrm{imp}}$ as
\begin{align}
  Z_{\mathrm{imp}}^{\kappa_{\sigma}}(t)&=\Tr\left[e^{\kappa_{\sigma}\sum_kV_k\hat{a}_{\sigma}^{\dag}\hat{c}_{\sigma,k}}\hat{\rho}_{\mathrm{full}}(t)\right]/\Tr\rhoop_{\mathrm{bath}}^{\mathrm{th}}
 \nonumber \\ 
 &= \int\mathcal{D}[\bar{\bm{a}},\bm{a}]
  \mathcal{K}[\bar{\bm{a}},\bm{a}]\mathcal{I}[\bar{\bm{a}},\bm{a}]
  \mathcal{Y}_{\kappa_{\sigma}}[\bar{\bm{a}}_{\sigma},\bm{a}_{\sigma}],
\end{align}
where 
\begin{align}
  \mathcal{Y}_{\kappa_{\sigma}}[\bar{\bm{a}}_{\sigma},\bm{a}_{\sigma}]=
  e^{-\im\kappa_{\sigma}\int_{\mathcal{C}}\dd\tau\bar{a}_{\sigma}(t^+)\Delta(t^+,\tau)a_{\sigma}(\tau)}.
\end{align}
Here $t^+$ denotes time $t$ on forward branch. Then the particle current can be written as
\begin{align}\label{eq:current2}
  \current_{\sigma}(t)&=2\Im\qty{Z^{-1}_{\mathrm{imp}}(t)\Tr_{\mathrm{imp}}\eval{\qty[\fdv{\kappa_{\sigma}}Z^{\kappa_{\sigma}}_{\mathrm{imp}}(t)]}_{\kappa_{\sigma}=0}} \nonumber \\ 
  &= -2\Re\int_{\mathcal{C}}\dd\tau\Delta(t^+,\tau)\expval*{\bar{a}_{\sigma}^+(t)a_{\sigma}(\tau)}.
\end{align}

To numerically evaluate the particle current, Eq. \eqref{eq:current2} should be discretized, similar to the PI itself. 
We use a similar discretization procedure to QuaPI, in which we treat $\bar{a}_{\sigma}^+(t)$ as a
segment rather than a single variable~\cite{chen2023-heat}:
\begin{align}
  \abar_{\sigma}^+(t)\to \frac{1}{\delta t}\int_{(M-\frac{1}{2})\delta t}^{(M+\frac{1}{2})\delta t}
  \abar_{\sigma,M}^+\dd t,
\end{align}
and discretize $\Delta(t^+,\tau)$ accordingly. Noticing that the time
of $\bar{a}_{\sigma}^+(t)$ is ahead of any other Grassmann variables
in normal time axis, and we arrive at the discretized expression of $\current_{\sigma}(t)$:
\begin{align}
  &\current_{\sigma}(t) \nonumber \\ 
  =& -\frac{2}{\delta t}\Re\qty[\sum_{k<M}\qty(\Delta^{++}_{Mk}\expval*{\bar{a}_{\sigma,M}^+a_{\sigma,k}^+}-
  \Delta^{+-}_{Mk}\expval*{\bar{a}_{\sigma,M}^+a_{\sigma,k}^-})].
\end{align}
In case of multiple baths, the total hybridization function becomes
$\Delta(\tau,\tau')=\sum_{\nu}\Delta^{\nu}(\tau,\tau')$, where
$\Delta^{\nu}$ is the hybridization functions for $\nu$-th
bath. Then the particle current $\current_{\sigma}^{\nu}(t)$ from the $\nu$-th bath can be calculated as
\begin{align}
  &\current_{\sigma}^{\nu}(t) \nonumber \\ 
  =&-\frac{2}{\delta t}\Re\qty[\sum_{k<M}\qty(\Delta^{\nu, ++}_{Mk}\expval*{\bar{a}_{\sigma,M}^+a_{\sigma,k}^+}-
  \Delta^{\nu, +-}_{Mk}\expval*{\bar{a}_{\sigma,M}^+a_{\sigma,k}^-})],
\end{align}
where $\Delta_{jk}^{\nu, \varsigma\varsigma'}$ are the discretized hybridization functions for $\Delta^{\nu}(\tau,\tau')$.

\section{Construction of $\mathcal{K}$}\label{app:K}

\begin{figure*}
  \includegraphics[width=2\columnwidth]{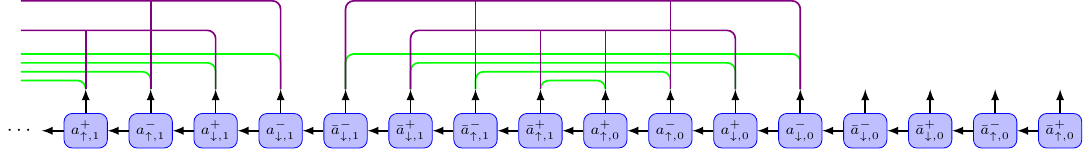} 
  \caption{Construction of the the bare impurity propagator as a Grassmann MPS by splitting $\gK$ into products of two-body and four-body terms. The green lines represent the two-body terms from the on site term of the impurity Hamiltonian, while the purple lines represent the four-body terms from the Coulomb interaction term of the impurity Hamiltonian.
    }
    \label{fig:fig1S}
\end{figure*}

The bare impurity dynamics $\gK$ of the PI can be generally constructed by
\begin{align}\label{eq:Kdis}
    \mathcal{K}[\bar{\bm{a}},\bm{a}]=&
    \mel*{-\bolda_M^+}{e^{-\im \Himp\delta t}}{\bolda_{M-1}^+}\cdots
    \mel*{\bolda_2^+}{e^{-\im \Himp\delta t}}{\bolda_1^+}\times \nonumber \\
    &\mel*{\bolda_{1}^+}{\hat{\rho}_{\mathrm{imp}}(0)}{\bolda_{1}^-} 
    \mel*{\bolda_1^-}{e^{\im \Himp\delta t}}{\bolda_{2}^-}\times\cdots \nonumber \\
    &\times\mel*{\bolda_{M-1}^-}{e^{\im \Himp\delta t}}{\bolda_M^-},
\end{align}
where $\bolda_k^{\pm}=\{a_{\uparrow,k}^{\pm},a_{\downarrow,k}^{\pm}\}$. For $\rhoop_{{\rm imp}}(0) = \vert 0\rangle\langle 0\vert$, we have
\begin{align}
\mel*{\bm{a}_{1}^+}{\rhoop_{\mathrm{imp}}(0)}{\bm{a}_{1}^-} = 1.
\end{align}
For the SIAM, each term in Eq. \eqref{eq:Kdis} is
\begin{align}
&\mel*{\bolda_k^{+}}{e^{-\im \Himp\delta t}}{\bolda_{k-1}^{+}} \nonumber \\
=& e^{g\sum_{\sigma} \abar_{\sigma, k}^+ a_{\sigma, k-1}^++g^2(e^{- \im \delta t U}-1)\abar_{\uparrow, k}^+ \abar_{\downarrow, k}^+ a_{\downarrow, k-1}^+ a_{\uparrow, k-1}^+}
\end{align}
for the forward branch with $g=e^{-\im(\epsilon_d - U/2)\delta t}$ and $\bar{g}$ being its complex conjugate, and 
\begin{align}
&\mel*{\bolda_{k-1}^{-}}{e^{\im \Himp\delta t}}{\bolda_{k}^{-}} \nonumber \\ 
=& e^{\bar{g} \sum_{\sigma} \abar_{\sigma, k-1}^- a_{\sigma, k}^- +\bar{g}^2(e^{ \im \delta t U}-1)\abar_{\uparrow, k-1}^- \abar_{\downarrow, k-1}^- a_{\downarrow, k}^- a_{\uparrow, k}^-}
\end{align}
for the backward branch. Thus $\gK$ can be constructed as a Grassmann MPS by products of two-body terms and four-body terms with forms $e^{\lambda\xi_i\xi_j}$ and $e^{\lambda\xi_i\xi_j\xi_k\xi_l}$ respectively as shown in Fig.~\ref{fig:fig1S}, where each term can be easily built as a Grassmann MPS and there are only $O(M)$ such terms. With our ordering of the Grassmann variables the computational cost of building $\gK$ is negligible compared to that of $\gI$, and $\gK$ can be built exactly with a moderate bond dimension $\chi_{\gK}<100$, which can be further lowered down to $\approx 20$ if MPS compression is performed with the same bond truncation tolerance used for $\gI$.

\section{Construction of $\mathcal{I}$}\label{app:I}

\begin{figure*}
  \includegraphics[width=2\columnwidth]{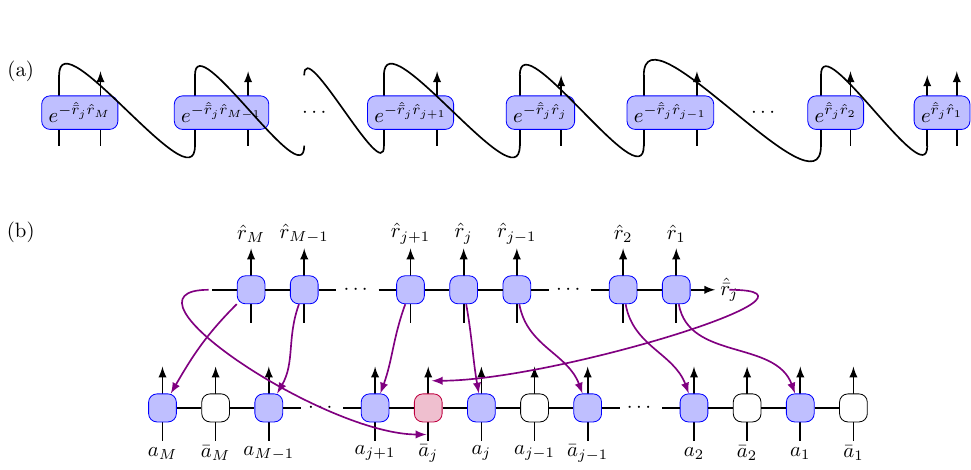} 
  \caption{(a) The coefficient tensor of the partial IF can be built by applying a series of bosonic two-body operators onto the vacuum state (the horizontal direction is the time direction). The whole operation is naturally an MPO of bond dimension $2$. (b) Moving the raising operators corresponding to $\bar{a}_j$ to the correct position, and inserting the identity operator (indicated by the white squares) in the places with no raising operators to form a global MPO with the same ordering as our predefined ordering of the Grassmann variables. Multiplying the resulting MPO onto the vacuum state (written as an MPS with bond dimension $1$) results in the Grassmann MPS for the partial IF.
    }
    \label{fig:fig2S}
\end{figure*}


Similar to the construction of $\mathcal{K}$, in principle we can split the IF as the product of $O(M^2)$ two-body terms in the form $e^{\lambda \xi_i\xi_j}$, and then multiply them together to obtain the MPS-IF. However with this approach there will be $O(M^2)$ Grassmann MPS multiplications, and one needs to perform an MPS compression after each of them (otherwise the bond dimension grows exponentially), which will be very inefficient.

Here we adopt a more efficient strategy, following the spirit of TEMPO~\cite{StrathearnLovett2018,strathearn2020-modelling,gribben2022-exact}, to construct the MPS-IF. 
For this we first define the partial IF as
\begin{align}\label{eq:partialIF1}
\gI^{\zeta}_{\sigma,j}[\bar{a}_{\sigma,j}^{\zeta},\bm{a}_{\sigma}] = e^{-\bar{a}_{\sigma,j}^{\zeta}\sum_{\zeta',k}\Delta^{\zeta\zeta'}_{j,k}a_{\sigma,k}^{\zeta'}},
\end{align}
with which we can rewrite the discretized IF as
\begin{align}
  \mathcal{I}_{\sigma}[\bar{\bm{a}}_{\sigma},\bm{a}_{\sigma}]=\prod_{j,\zeta}\mathcal{I}_{\sigma,j}^{\zeta}[\bar{a}_{\sigma,j}^{\zeta},\bm{a}_{\sigma}].
\end{align}
We will compactly write each partial IF as a Grassmann MPS and then obtain the MPS-IF using $O(M)$ Grassmann MPS multiplications.
To write the partial IF as a single MPS, we expand Eq. \eqref{eq:partialIF1} as
\begin{align}\label{eq:partialIF2}
\gI^{\zeta}_{\sigma,j}[\bar{a}_{\sigma,j}^{\zeta},\bm{a}_{\sigma}] = 1 - \sum_{\zeta', k}\abar^{\zeta}_{\sigma, j} \Delta^{\zeta\zeta'}_{j,k} a^{\zeta'}_{\sigma, k}, 
\end{align}
where higher order terms vanish because the quadratic terms in the exponent share the same Grassmann variable $\abar^{\zeta}_{\sigma, j}$. 
Now we note that the Grassmann vacuum (e.g., the scalar $1$) for a Grassmann space of $n$ Grassmann variables from $\xi_1$ to $\xi_n$, can be written as a Grassmann tensor $\gV$ such that the only nonvanishing element of its coefficient tensor is $V^{0,0,0,\cdots, 0}=1$. We can see that the effect of any quadratic term $\xi_i\xi_j$ (with $i>j$) on $\gV$ is
\begin{align}
\xi_i\xi_j\gV = \sum_{\bm{i}} \left(\hat{r}_i\hat{r}_j V^{i_n, \cdots, i_1}\right) \xi^{i_n}_n \xi^{i_{n-1}}_{n-1} \cdots \xi^{i_1}_1,
\end{align}
where
\begin{align}
\hat{r}_i = \left[\begin{array}{cc} 1 & 0 \\ 0 & 0\end{array}\right]
\end{align}
is the bosonic raising operator on site $i$, 
namely the multiplication of a quadratic term $\xi_i\xi_j$ on a Grassmann vacuum boils down to applying a bosonic Pauli string $\hat{r}_i \hat{r}_j$ on into the coefficient tensor (or MPS).
In this case no additional sign appears between $\hat{r}_i$ and $\hat{r}_j$ since the operand is vacuum. As a result, the rhs of Eq. \eqref{eq:partialIF2} will be equivalent to applying a series of bosonic operator in the form $1 - \sum_{i, j}\lambda_{ij}\hat{r}_i\hat{r}_j $ on the coefficient tensor of $\gV$. The sum of the bosonic terms can also be restored into the product form as $e^{-\sum_{i,j}\lambda_{ij}\hat{r}_i\hat{r}_j}$ since $e^{-\lambda_{ij}\hat{r}_i\hat{r}_j} = 1 - \lambda_{ij}\hat{r}_i \hat{r}_j$. With these observations, the partial IF can be built as a Grassmann MPS as follows.
We first apply the bosonic product operator $e^{-\sum_{i,j}\lambda_{ij}\hat{r}_i\hat{r}_j}$ corresponding to Eq. \eqref{eq:partialIF2} onto the coefficient tensor of vacuum (which is written as an MPS with bond dimension $1$), as shown in Fig.~\ref{fig:fig2S}(a). Since the two-body terms in Fig.~\ref{fig:fig2S}(a) all share one same index, it is naturally a matrix product operator (MPO) of bond dimension $2$ (which is because that the rank of the two-body operator $e^{\hat{r}_i \hat{r}_j}$ is $2$). Since the shared index $\hat{\bar{r}}_j$ corresponds to the Grassmann variable $\abar^{\zeta}_{\sigma, j}$, we still need to move it to the correct position to force the MPO to be of the same ordering as our predefined ordering for the Grassmann variables, which can be done using a series of bosonic swap operations as shown in Fig.~\ref{fig:fig2S}(b). In practice we observe that these swap operations generally do not increase the bond dimension of the MPO, even with a very low bond truncation tolerance as $10^{-10}$. By applying the resulting MPO onto the vacuum state, we obtain the Grassmann MPS for the partial IF in Eq. \eqref{eq:partialIF1} with bond dimension $2$.

\section{Multiplication of two site tensors}\label{app:sitemult}

\begin{figure}
  \includegraphics[]{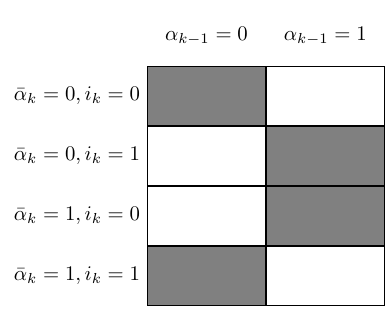} 
  \caption{Storage of a rank-$3$ site tensor as a block sparse matrix. Only the gray blocks, which satisfy the even parity condition, can be nonzero.
    }
    \label{fig:fig3S}
\end{figure}

The Grassmann MPS is naturally implemented as a $Z_2$-symmetric MPS. In numerical implementation, the $k$-th site tensor $A^{i_k}_{\bar{\alpha}_k, \alpha_{k-1}}$ of a Grassmann MPS $\gA$ can be stored as in Fig.~\ref{fig:fig3S}. In total, there are $8$ possible combinations for the three indices $i_k$, $\bar{\alpha}_k$ and $\alpha_{k-1}$ since each of them can be $0$ or $1$. However, due to the \textit{even parity condition} as shown in the main text, there are only four combinations which can be nonzero, as indicated by the $4$ gray blocks in Fig.~\ref{fig:fig3S}. This property can be used to save half of the storage by only storing the nonzero blocks,
similar to the case of general Abelian-symmetric MPSs~\cite{SinghVidal2011}.

With our definition of Grassmann MPS, the multiplication of two Grassmann MPSs can be performed site by site, and the central subroutine to implement is the multiplication of two site tensors, for which the anticommutative nature of the Grassmann variables has to be explicitly taken into account.
The first step for the multiplication of the two site tensors $A^{i_k}_{\bar{\alpha}_k, \alpha_{k-1}}$ and $B^{i_k'}_{\bar{\alpha}_k', \alpha_{k-1}'}$ is a direct tensor product with proper reordering of the local Grassmann variables as
\begin{align}\label{eq:tensorproduct}
& \left(\sum_{\bar{\alpha}_k, i_k, \alpha_{k-1}} A^{i_k}_{\bar{\alpha}_k, \alpha_{k-1}} \etabar_k^{\bar{\alpha}_k}\xi_k^{i_k} \eta_{k-1}^{\alpha_{k-1}}\right) \nonumber \\ 
&\times \left(\sum_{\bar{\alpha}_k', i_k', \alpha_{k-1}'} B^{i_k'}_{\bar{\alpha}_k', \alpha_{k-1}'} (\etabar_k')^{\bar{\alpha}_k'}\xi_k^{i_k'} (\eta_{k-1}')^{\alpha_{k-1}'} \right) \nonumber \\
= & \sum_{\bar{\alpha}_k, i_k, \alpha_{k-1}, \bar{\alpha}_k', i_k', \alpha_{k-1}'} (-1)^{\alpha_{k-1}i_k'} A^{i_k}_{\bar{\alpha}_k, \alpha_{k-1}} B^{i_k'}_{\bar{\alpha}_k', \alpha_{k-1}'} 
\times \nonumber \\ &(\etabar_k')^{\bar{\alpha}_k'} 
\etabar_k^{\bar{\alpha}_k} \xi_k^{i_k} \xi_k^{i_k'} \eta_{k-1}^{\alpha_{k-1}} (\eta_{k-1}')^{\alpha_{k-1}'},
\end{align}
where in the second line we have moved the Grassmann variable $\eta_k'$ from the $4$-th position to the first position, for which no minus sign occurs since the even parity of the first $3$ Grassmann variables, and then moved $\xi_k^{i_k'}$ to the $4$-th position, which induces a sign $(-1)^{\alpha_{k-1}i_k'}$ due to the interchange with $ \eta_{k-1}^{\alpha_{k-1}}$.

The second step is to regroup the indices to restore the form of a rank-$3$ site tensor, for this we first denote the rank-$6$ coefficient tensor in Eq. \eqref{eq:tensorproduct} as 
\begin{align}\label{eq:W}
W_{\bar{\alpha}_k',\bar{\alpha}_k, \alpha_{k-1}, \alpha_{k-1}'}^{i_k, i_k'} = (-1)^{\alpha_{k-1}i_k'} A^{i_k}_{\bar{\alpha}_k, \alpha_{k-1}} B^{i_k'}_{\bar{\alpha}_k', \alpha_{k-1}'}.
\end{align}
Since the physical Grassmann variables of the two input site tensor are the same, we simply multiply them together, which results in a rank-$5$ Grassmann tensor denoted as
\begin{align}\label{eq:V}
\sum_{\bar{\alpha}_k, i_k, \alpha_{k-1}, \bar{\alpha}_k', \alpha_{k-1}'} &V_{\bar{\alpha}_k',\bar{\alpha}_k, \alpha_{k-1}, \alpha_{k-1}'}^{i_k} \times \nonumber \\ 
&(\etabar_k')^{\bar{\alpha}_k'} \etabar_k^{\bar{\alpha}_k}\xi_k^{i_k} \eta_{k-1}^{\alpha_{k-1}} (\eta_{k-1}')^{\alpha_{k-1}'},
\end{align}
where $V$ is related to $W$ in Eq. \eqref{eq:W} as
\begin{align}\label{eq:GVmult}
V_{\bar{\alpha}_k',\bar{\alpha}_k, \alpha_{k-1}, \alpha_{k-1}'}^{0} &= W_{\bar{\alpha}_k',\bar{\alpha}_k, \alpha_{k-1}, \alpha_{k-1}'}^{0, 0}; \\
V_{\bar{\alpha}_k',\bar{\alpha}_k, \alpha_{k-1}, \alpha_{k-1}'}^{1} &= W_{\bar{\alpha}_k',\bar{\alpha}_k, \alpha_{k-1}, \alpha_{k-1}'}^{1, 0} + W_{\bar{\alpha}_k',\bar{\alpha}_k, \alpha_{k-1}, \alpha_{k-1}'}^{0, 1}.
\end{align}
Then we perform the fusion operation of two auxiliary $Z_2$-symmetric tensor indices in the same way as the bosonic case [there will be no sign issue since we have already placed the corresponding Grassmann variables together in Eq. \eqref{eq:tensorproduct}]. For example, the fusion of the two auxiliary indices $\bar{\alpha}_k'$ and $\bar{\alpha}_k$ will result a rank-$4$ Grassmann tensor denoted as
\begin{align}\label{eq:C}
\sum_{\bar{\alpha}_k'', i_k, \alpha_{k-1}, \alpha_{k-1}'} C^{i_k}_{\bar{\alpha}_k'', \alpha_{k-1}, \alpha_{k-1}'} (\etabar_k'')^{\bar{\alpha}_k''} \xi_k^{i_k}  \eta_{k-1}^{\alpha_{k-1}} (\eta_{k-1}')^{\alpha_{k-1}'}.
\end{align}
The relation between Eqs. \eqref{eq:V} and \eqref{eq:C} is as follows. First, the two auxiliary Grassmann variables $\etabar_k'$ and $\etabar_k$ are fused into one Grassmann variable $\etabar_k''$ as
\begin{align}
(\etabar_k')^0 \etabar_k^0,  (\etabar_k')^1 \etabar_k^1 \rightarrow (\etabar_k'')^0; \\
(\etabar_k')^1 \etabar_k^0,  (\etabar_k')^0 \etabar_k^1 \rightarrow (\etabar_k'')^1.
\end{align}
Correspondingly, the auxiliary indices of the coefficient tensor $V$ are fused together as
\begin{align}
C^{i_k}_{0, \alpha_{k-1}, \alpha_{k-1}'} = V_{(0,0), \alpha_{k-1}, \alpha_{k-1}'}^{i_k} \oplus_1 V_{(1,1), \alpha_{k-1}, \alpha_{k-1}'}^{i_k}; \\
C^{i_k}_{1, \alpha_{k-1}, \alpha_{k-1}'} = V_{(1,0), \alpha_{k-1}, \alpha_{k-1}'}^{i_k} \oplus_1 V_{(0,1), \alpha_{k-1}, \alpha_{k-1}'}^{i_k}.
\end{align}
Here $(x, y)$ means to group the two indices $x, y$ of a dense array into a single index (the ``reshape'' operation), and $X\oplus_1 Y$ means the direct stacking of the first dimension of the two input dense tensors $X$ and $Y$. By fusing the two auxiliary indices $\alpha_{k-1}$ and $\alpha_{k-1}'$ in a similar way, we can finally obtain a rank-$3$ site tensor with local basis of Grassmann variables $(\etabar_k'')^{\bar{\alpha}_k''} \xi_k^{i_k} (\eta_{k-1}'')^{\alpha_{k-1}''} $.

\section{Integration of Grassmann MPS}\label{app:integration}

\begin{figure*}
  \includegraphics[width=2\columnwidth]{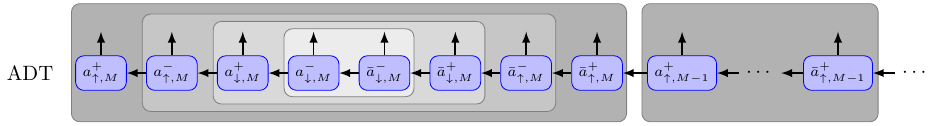} 
  \caption{Integration of pair-wise physical Grassmann variables based on the ADT, where the integration is performed from inner to outer within each time step.
    }
    \label{fig:fig4S}
\end{figure*}

Computing observables, such as the impurity Green's functions, requires to integrate the Grassmann MPS to obtain ordinary scalars. The most straightforward way to perform integration of the Grassmann MPS is to first obtain the augmented density tensor (ADT) $\gA[\boldabar, \bolda] = \gK[\boldabar, \bolda] \sum_{\sigma}\gI_{\sigma}[\boldabar_{\sigma}, \bolda_{\sigma}]$, as done in TEMPO~\cite{StrathearnLovett2018}, and then perform integration based on $\gA$, which can be done straightforwardly as in Fig.~\ref{fig:fig4S} for our specific ordering of the Grassmann variables. Mathematically, the integration of a nearest-neighbour pair of physical Grassmann variables $\xi_{k+1}$ and $\xi_k$ with $\xi_k = \bar{\xi}_{k+1}$ can be calculated as
\begin{align}
&\int \dd\xi_k\dd\xi_{k+1}e^{-\xi_k\xi_{k+1}} \int \dd\etabar_k\dd \eta_k e^{-\etabar_k\eta_k} \nonumber \\ 
&\left(\sum_{i_{k+1}, \bar{\alpha}_{k+1}, \alpha_k } A^{i_{k+1}}_{\bar{\alpha}_{k+1}, \alpha_k} \etabar_{k+1}^{\bar{\alpha}_{k+1}} \xi_{k+1}^{i_{k+1}} \eta_{k}^{\alpha_{k}}\right) \times \nonumber \\ &\left(\sum_{i_{k}, \bar{\alpha}_{k}, \alpha_{k-1} } A^{i_{k}}_{\bar{\alpha}_{k}, \alpha_{k-1}}  \etabar_{k}^{\bar{\alpha}_{k}}\xi_{k}^{i_k} \eta_{k-1}^{\alpha_{k-1}}\right) \nonumber \\ 
=& \sum_{\bar{\alpha}_{k+1}, \alpha_{k-1}} \left(\sum_{i_k, \alpha_k}  A^{i_{k}}_{\bar{\alpha}_{k+1}, \alpha_k} A^{i_{k}}_{\alpha_{k}, \alpha_{k-1}} \right) \etabar_{k+1}^{\bar{\alpha}_{k+1}} \eta_{k-1}^{\alpha_{k-1}}.
\end{align}
The physical Grassmann variables at each time step are integrated from inner to outer in Fig.~\ref{fig:fig4S}, 
with the overall computational cost scales as $O(M\chi_{\gA}^3)$.

\begin{figure*}
  \includegraphics[width=2\columnwidth]{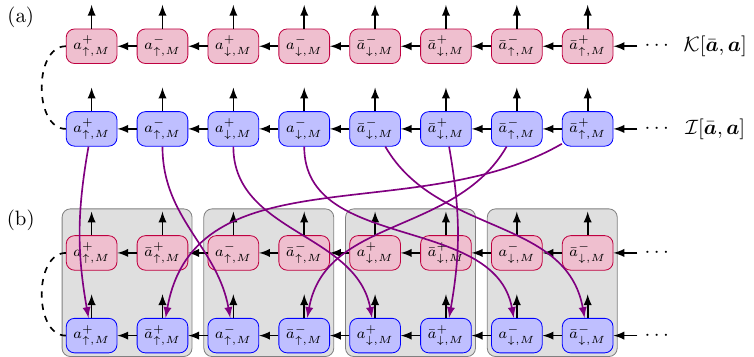} 
  \caption{(a) The original ordering of the Grassmann variables for $\gK$ and $\gI$. (b) Reordering of the local Grassmann variables at each time step, and then integrating the pair-wise physical Grassmann variables from left to right, where the multiplication of the same physical Grassmann variables from $\gK$ and $\gI$ are performed on the fly. 
    }
    \label{fig:fig5S}
\end{figure*}

Instead of building the ADT explicitly and then compute observables based on the ADT, we directly compute observables based on $\gK$ and $\gI_{\sigma}$. Except the saving of the computational cost to perform the multiplication of $\gK$ and $\gI_{\sigma}$, the calculation of the observables can also be performed more efficiently, as demonstrated in Fig.~\ref{fig:fig5S} for spinless fermions (in which case we use $\gI$ instead of $\gI_{\sigma}$). In this approach, we first perform a series of local swap operations on $\gK$ and $\gI$ such that the ordering of the Grassmann variables within each time step is changed from Fig.~\ref{fig:fig5S}(a) to Fig.~\ref{fig:fig5S}(b). Then we integrate the physical Grassmann variables from left to right as shown in Fig.~\ref{fig:fig5S}(b).
The overall computational cost of this zipup algorithm is thus $O(M(\chi_{\gK}^3 + \chi^3 + \chi_{\gK}\chi^2 + \chi_{\gK}^2\chi ))$, where the first two terms are due to the local swap operations performed within $\gK$ and $\gI$, and the last two terms are due to the tensor contractions performed inside each block of Fig.~\ref{fig:fig5S}(b). For SIAM we observe that $\chi_{\gK}=32$ for nonzero $U$, and $\chi_{\gK} \approx 20$ if we compress it with the same bond truncation tolerance as used for $\gI_{\sigma}$. Therefore the terms in the computational cost which contain $\chi_{\gK}$ is simply ignored. 

The zipup algorithm can be straightforwardly generalized for more than two GMPSs.
In fact, for SIAM with nonzero $U$, we use three GMPSs, namely $\gK$, $\gI_{\uparrow}$ and $\gI_{\downarrow}$ for the integration. Similarly, the computational cost for computing observables scales as $O(M(\chi_{\gK}^3 + 2\chi^3  + \chi_{\gK}^2\chi^2 + 2\chi_{\gK}\chi^3 ))$ (the expression in the main text is obtained by neglecting the first three terms), and the memory cost is $O(\chi_{\gK}\chi^2)$, which is the size of the largest intermediate tensor appears during the integration.


\begin{thebibliography}{59}%
\makeatletter
\providecommand \@ifxundefined [1]{%
 \@ifx{#1\undefined}
}%
\providecommand \@ifnum [1]{%
 \ifnum #1\expandafter \@firstoftwo
 \else \expandafter \@secondoftwo
 \fi
}%
\providecommand \@ifx [1]{%
 \ifx #1\expandafter \@firstoftwo
 \else \expandafter \@secondoftwo
 \fi
}%
\providecommand \natexlab [1]{#1}%
\providecommand \enquote  [1]{``#1''}%
\providecommand \bibnamefont  [1]{#1}%
\providecommand \bibfnamefont [1]{#1}%
\providecommand \citenamefont [1]{#1}%
\providecommand \href@noop [0]{\@secondoftwo}%
\providecommand \href [0]{\begingroup \@sanitize@url \@href}%
\providecommand \@href[1]{\@@startlink{#1}\@@href}%
\providecommand \@@href[1]{\endgroup#1\@@endlink}%
\providecommand \@sanitize@url [0]{\catcode `\\12\catcode `\$12\catcode
  `\&12\catcode `\#12\catcode `\^12\catcode `\_12\catcode `\%12\relax}%
\providecommand \@@startlink[1]{}%
\providecommand \@@endlink[0]{}%
\providecommand \url  [0]{\begingroup\@sanitize@url \@url }%
\providecommand \@url [1]{\endgroup\@href {#1}{\urlprefix }}%
\providecommand \urlprefix  [0]{URL }%
\providecommand \Eprint [0]{\href }%
\providecommand \doibase [0]{https://doi.org/}%
\providecommand \selectlanguage [0]{\@gobble}%
\providecommand \bibinfo  [0]{\@secondoftwo}%
\providecommand \bibfield  [0]{\@secondoftwo}%
\providecommand \translation [1]{[#1]}%
\providecommand \BibitemOpen [0]{}%
\providecommand \bibitemStop [0]{}%
\providecommand \bibitemNoStop [0]{.\EOS\space}%
\providecommand \EOS [0]{\spacefactor3000\relax}%
\providecommand \BibitemShut  [1]{\csname bibitem#1\endcsname}%
\let\auto@bib@innerbib\@empty
\bibitem [{\citenamefont {Anderson}(1961)}]{anderson1961-localized}%
  \BibitemOpen
  \bibfield  {author} {\bibinfo {author} {\bibfnamefont {P.~W.}\ \bibnamefont
  {Anderson}},\ }\bibfield  {title} {\bibinfo {title} {Localized magnetic
  states in metals},\ }\href {https://doi.org/10.1103/physrev.124.41}
  {\bibfield  {journal} {\bibinfo  {journal} {Phys. Rev.}\ }\textbf {\bibinfo
  {volume} {124}},\ \bibinfo {pages} {41} (\bibinfo {year} {1961})}\BibitemShut
  {NoStop}%
\bibitem [{\citenamefont {Georges}\ \emph {et~al.}(1996)\citenamefont
  {Georges}, \citenamefont {Kotliar}, \citenamefont {Krauth},\ and\
  \citenamefont {Rozenberg}}]{GeorgesRozenberg1996}%
  \BibitemOpen
  \bibfield  {author} {\bibinfo {author} {\bibfnamefont {A.}~\bibnamefont
  {Georges}}, \bibinfo {author} {\bibfnamefont {G.}~\bibnamefont {Kotliar}},
  \bibinfo {author} {\bibfnamefont {W.}~\bibnamefont {Krauth}},\ and\ \bibinfo
  {author} {\bibfnamefont {M.~J.}\ \bibnamefont {Rozenberg}},\ }\bibfield
  {title} {\bibinfo {title} {Dynamical mean-field theory of strongly correlated
  fermion systems and the limit of infinite dimensions},\ }\href
  {https://doi.org/10.1103/RevModPhys.68.13} {\bibfield  {journal} {\bibinfo
  {journal} {Rev. Mod. Phys.}\ }\textbf {\bibinfo {volume} {68}},\ \bibinfo
  {pages} {13} (\bibinfo {year} {1996})}\BibitemShut {NoStop}%
\bibitem [{\citenamefont {Gull}\ \emph {et~al.}(2011)\citenamefont {Gull},
  \citenamefont {Millis}, \citenamefont {Lichtenstein}, \citenamefont
  {Rubtsov}, \citenamefont {Troyer},\ and\ \citenamefont
  {Werner}}]{GullWerner2011}%
  \BibitemOpen
  \bibfield  {author} {\bibinfo {author} {\bibfnamefont {E.}~\bibnamefont
  {Gull}}, \bibinfo {author} {\bibfnamefont {A.~J.}\ \bibnamefont {Millis}},
  \bibinfo {author} {\bibfnamefont {A.~I.}\ \bibnamefont {Lichtenstein}},
  \bibinfo {author} {\bibfnamefont {A.~N.}\ \bibnamefont {Rubtsov}}, \bibinfo
  {author} {\bibfnamefont {M.}~\bibnamefont {Troyer}},\ and\ \bibinfo {author}
  {\bibfnamefont {P.}~\bibnamefont {Werner}},\ }\bibfield  {title} {\bibinfo
  {title} {Continuous-time monte carlo methods for quantum impurity models},\
  }\href {https://doi.org/10.1103/RevModPhys.83.349} {\bibfield  {journal}
  {\bibinfo  {journal} {Rev. Mod. Phys.}\ }\textbf {\bibinfo {volume} {83}},\
  \bibinfo {pages} {349} (\bibinfo {year} {2011})}\BibitemShut {NoStop}%
\bibitem [{\citenamefont {Rubtsov}\ and\ \citenamefont
  {Lichtenstein}(2004)}]{RubtsovLichtenstein2004}%
  \BibitemOpen
  \bibfield  {author} {\bibinfo {author} {\bibfnamefont {A.~N.}\ \bibnamefont
  {Rubtsov}}\ and\ \bibinfo {author} {\bibfnamefont {A.~I.}\ \bibnamefont
  {Lichtenstein}},\ }\bibfield  {title} {\bibinfo {title} {Continuous-time
  quantum monte carlo method for fermions: Beyond auxiliary field framework},\
  }\href {https://doi.org/10.1134/1.1800216} {\bibfield  {journal} {\bibinfo
  {journal} {J. Exp. Theor. Phys. Lett.}\ }\textbf {\bibinfo {volume} {80}},\
  \bibinfo {pages} {61} (\bibinfo {year} {2004})}\BibitemShut {NoStop}%
\bibitem [{\citenamefont {Rubtsov}\ \emph {et~al.}(2005)\citenamefont
  {Rubtsov}, \citenamefont {Savkin},\ and\ \citenamefont
  {Lichtenstein}}]{RubtsovLichtenstein2005}%
  \BibitemOpen
  \bibfield  {author} {\bibinfo {author} {\bibfnamefont {A.~N.}\ \bibnamefont
  {Rubtsov}}, \bibinfo {author} {\bibfnamefont {V.~V.}\ \bibnamefont
  {Savkin}},\ and\ \bibinfo {author} {\bibfnamefont {A.~I.}\ \bibnamefont
  {Lichtenstein}},\ }\bibfield  {title} {\bibinfo {title} {Continuous-time
  quantum monte carlo method for fermions},\ }\href
  {https://doi.org/10.1103/PhysRevB.72.035122} {\bibfield  {journal} {\bibinfo
  {journal} {Phys. Rev. B}\ }\textbf {\bibinfo {volume} {72}},\ \bibinfo
  {pages} {035122} (\bibinfo {year} {2005})}\BibitemShut {NoStop}%
\bibitem [{\citenamefont {Werner}\ \emph {et~al.}(2006)\citenamefont {Werner},
  \citenamefont {Comanac}, \citenamefont {de' Medici}, \citenamefont {Troyer},\
  and\ \citenamefont {Millis}}]{WernerMillis2006}%
  \BibitemOpen
  \bibfield  {author} {\bibinfo {author} {\bibfnamefont {P.}~\bibnamefont
  {Werner}}, \bibinfo {author} {\bibfnamefont {A.}~\bibnamefont {Comanac}},
  \bibinfo {author} {\bibfnamefont {L.}~\bibnamefont {de' Medici}}, \bibinfo
  {author} {\bibfnamefont {M.}~\bibnamefont {Troyer}},\ and\ \bibinfo {author}
  {\bibfnamefont {A.~J.}\ \bibnamefont {Millis}},\ }\bibfield  {title}
  {\bibinfo {title} {Continuous-time solver for quantum impurity models},\
  }\href {https://doi.org/10.1103/PhysRevLett.97.076405} {\bibfield  {journal}
  {\bibinfo  {journal} {Phys. Rev. Lett.}\ }\textbf {\bibinfo {volume} {97}},\
  \bibinfo {pages} {076405} (\bibinfo {year} {2006})}\BibitemShut {NoStop}%
\bibitem [{\citenamefont {Gull}\ \emph {et~al.}(2008)\citenamefont {Gull},
  \citenamefont {Werner}, \citenamefont {Parcollet},\ and\ \citenamefont
  {Troyer}}]{GullTroyer2008}%
  \BibitemOpen
  \bibfield  {author} {\bibinfo {author} {\bibfnamefont {E.}~\bibnamefont
  {Gull}}, \bibinfo {author} {\bibfnamefont {P.}~\bibnamefont {Werner}},
  \bibinfo {author} {\bibfnamefont {O.}~\bibnamefont {Parcollet}},\ and\
  \bibinfo {author} {\bibfnamefont {M.}~\bibnamefont {Troyer}},\ }\bibfield
  {title} {\bibinfo {title} {Continuous-time auxiliary-field monte carlo for
  quantum impurity models},\ }\href
  {https://doi.org/10.1209/0295-5075/82/57003} {\bibfield  {journal} {\bibinfo
  {journal} {EPL}\ }\textbf {\bibinfo {volume} {82}},\ \bibinfo {pages} {57003}
  (\bibinfo {year} {2008})}\BibitemShut {NoStop}%
\bibitem [{\citenamefont {Werner}\ and\ \citenamefont
  {Millis}(2006)}]{WernerMillis2006b}%
  \BibitemOpen
  \bibfield  {author} {\bibinfo {author} {\bibfnamefont {P.}~\bibnamefont
  {Werner}}\ and\ \bibinfo {author} {\bibfnamefont {A.~J.}\ \bibnamefont
  {Millis}},\ }\bibfield  {title} {\bibinfo {title} {Hybridization expansion
  impurity solver: General formulation and application to kondo lattice and
  two-orbital models},\ }\href {https://doi.org/10.1103/PhysRevB.74.155107}
  {\bibfield  {journal} {\bibinfo  {journal} {Phys. Rev. B}\ }\textbf {\bibinfo
  {volume} {74}},\ \bibinfo {pages} {155107} (\bibinfo {year}
  {2006})}\BibitemShut {NoStop}%
\bibitem [{\citenamefont {Haule}(2007)}]{Haule2007}%
  \BibitemOpen
  \bibfield  {author} {\bibinfo {author} {\bibfnamefont {K.}~\bibnamefont
  {Haule}},\ }\bibfield  {title} {\bibinfo {title} {Quantum monte carlo
  impurity solver for cluster dynamical mean-field theory and electronic
  structure calculations with adjustable cluster base},\ }\href
  {https://doi.org/10.1103/PhysRevB.75.155113} {\bibfield  {journal} {\bibinfo
  {journal} {Phys. Rev. B}\ }\textbf {\bibinfo {volume} {75}},\ \bibinfo
  {pages} {155113} (\bibinfo {year} {2007})}\BibitemShut {NoStop}%
\bibitem [{\citenamefont {Werner}\ and\ \citenamefont
  {Millis}(2007)}]{WernerMillis2007}%
  \BibitemOpen
  \bibfield  {author} {\bibinfo {author} {\bibfnamefont {P.}~\bibnamefont
  {Werner}}\ and\ \bibinfo {author} {\bibfnamefont {A.~J.}\ \bibnamefont
  {Millis}},\ }\bibfield  {title} {\bibinfo {title} {High-spin to low-spin and
  orbital polarization transitions in multiorbital mott systems},\ }\href
  {https://doi.org/10.1103/PhysRevLett.99.126405} {\bibfield  {journal}
  {\bibinfo  {journal} {Phys. Rev. Lett.}\ }\textbf {\bibinfo {volume} {99}},\
  \bibinfo {pages} {126405} (\bibinfo {year} {2007})}\BibitemShut {NoStop}%
\bibitem [{\citenamefont {Werner}\ \emph {et~al.}(2008)\citenamefont {Werner},
  \citenamefont {Gull}, \citenamefont {Troyer},\ and\ \citenamefont
  {Millis}}]{WernerMillis2008}%
  \BibitemOpen
  \bibfield  {author} {\bibinfo {author} {\bibfnamefont {P.}~\bibnamefont
  {Werner}}, \bibinfo {author} {\bibfnamefont {E.}~\bibnamefont {Gull}},
  \bibinfo {author} {\bibfnamefont {M.}~\bibnamefont {Troyer}},\ and\ \bibinfo
  {author} {\bibfnamefont {A.~J.}\ \bibnamefont {Millis}},\ }\bibfield  {title}
  {\bibinfo {title} {Spin freezing transition and non-fermi-liquid self-energy
  in a three-orbital model},\ }\href
  {https://doi.org/10.1103/PhysRevLett.101.166405} {\bibfield  {journal}
  {\bibinfo  {journal} {Phys. Rev. Lett.}\ }\textbf {\bibinfo {volume} {101}},\
  \bibinfo {pages} {166405} (\bibinfo {year} {2008})}\BibitemShut {NoStop}%
\bibitem [{\citenamefont {Chan}\ \emph {et~al.}(2009)\citenamefont {Chan},
  \citenamefont {Werner},\ and\ \citenamefont {Millis}}]{ChanMillis2009}%
  \BibitemOpen
  \bibfield  {author} {\bibinfo {author} {\bibfnamefont {C.-K.}\ \bibnamefont
  {Chan}}, \bibinfo {author} {\bibfnamefont {P.}~\bibnamefont {Werner}},\ and\
  \bibinfo {author} {\bibfnamefont {A.~J.}\ \bibnamefont {Millis}},\ }\bibfield
   {title} {\bibinfo {title} {Magnetism and orbital ordering in an interacting
  three-band model: A dynamical mean-field study},\ }\href
  {https://doi.org/10.1103/PhysRevB.80.235114} {\bibfield  {journal} {\bibinfo
  {journal} {Phys. Rev. B}\ }\textbf {\bibinfo {volume} {80}},\ \bibinfo
  {pages} {235114} (\bibinfo {year} {2009})}\BibitemShut {NoStop}%
\bibitem [{\citenamefont {Aoki}\ \emph {et~al.}(2014)\citenamefont {Aoki},
  \citenamefont {Tsuji}, \citenamefont {Eckstein}, \citenamefont {Kollar},
  \citenamefont {Oka},\ and\ \citenamefont {Werner}}]{AokiWerner2014}%
  \BibitemOpen
  \bibfield  {author} {\bibinfo {author} {\bibfnamefont {H.}~\bibnamefont
  {Aoki}}, \bibinfo {author} {\bibfnamefont {N.}~\bibnamefont {Tsuji}},
  \bibinfo {author} {\bibfnamefont {M.}~\bibnamefont {Eckstein}}, \bibinfo
  {author} {\bibfnamefont {M.}~\bibnamefont {Kollar}}, \bibinfo {author}
  {\bibfnamefont {T.}~\bibnamefont {Oka}},\ and\ \bibinfo {author}
  {\bibfnamefont {P.}~\bibnamefont {Werner}},\ }\bibfield  {title} {\bibinfo
  {title} {Nonequilibrium dynamical mean-field theory and its applications},\
  }\href {https://doi.org/10.1103/RevModPhys.86.779} {\bibfield  {journal}
  {\bibinfo  {journal} {Rev. Mod. Phys.}\ }\textbf {\bibinfo {volume} {86}},\
  \bibinfo {pages} {779} (\bibinfo {year} {2014})}\BibitemShut {NoStop}%
\bibitem [{\citenamefont {Troyer}\ and\ \citenamefont
  {Wiese}(2005)}]{TroyerWiese2005}%
  \BibitemOpen
  \bibfield  {author} {\bibinfo {author} {\bibfnamefont {M.}~\bibnamefont
  {Troyer}}\ and\ \bibinfo {author} {\bibfnamefont {U.-J.}\ \bibnamefont
  {Wiese}},\ }\bibfield  {title} {\bibinfo {title} {Computational complexity
  and fundamental limitations to fermionic quantum monte carlo simulations},\
  }\href {https://doi.org/10.1103/PhysRevLett.94.170201} {\bibfield  {journal}
  {\bibinfo  {journal} {Phys. Rev. Lett.}\ }\textbf {\bibinfo {volume} {94}},\
  \bibinfo {pages} {170201} (\bibinfo {year} {2005})}\BibitemShut {NoStop}%
\bibitem [{\citenamefont {Wolf}\ \emph {et~al.}(2015)\citenamefont {Wolf},
  \citenamefont {Go}, \citenamefont {McCulloch}, \citenamefont {Millis},\ and\
  \citenamefont {Schollw\"ock}}]{WolfSchollwock2015}%
  \BibitemOpen
  \bibfield  {author} {\bibinfo {author} {\bibfnamefont {F.~A.}\ \bibnamefont
  {Wolf}}, \bibinfo {author} {\bibfnamefont {A.}~\bibnamefont {Go}}, \bibinfo
  {author} {\bibfnamefont {I.~P.}\ \bibnamefont {McCulloch}}, \bibinfo {author}
  {\bibfnamefont {A.~J.}\ \bibnamefont {Millis}},\ and\ \bibinfo {author}
  {\bibfnamefont {U.}~\bibnamefont {Schollw\"ock}},\ }\bibfield  {title}
  {\bibinfo {title} {Imaginary-time matrix product state impurity solver for
  dynamical mean-field theory},\ }\href
  {https://doi.org/10.1103/PhysRevX.5.041032} {\bibfield  {journal} {\bibinfo
  {journal} {Phys. Rev. X}\ }\textbf {\bibinfo {volume} {5}},\ \bibinfo {pages}
  {041032} (\bibinfo {year} {2015})}\BibitemShut {NoStop}%
\bibitem [{\citenamefont {Fei}\ \emph {et~al.}(2021)\citenamefont {Fei},
  \citenamefont {Yeh},\ and\ \citenamefont {Gull}}]{FeiGull2021}%
  \BibitemOpen
  \bibfield  {author} {\bibinfo {author} {\bibfnamefont {J.}~\bibnamefont
  {Fei}}, \bibinfo {author} {\bibfnamefont {C.-N.}\ \bibnamefont {Yeh}},\ and\
  \bibinfo {author} {\bibfnamefont {E.}~\bibnamefont {Gull}},\ }\bibfield
  {title} {\bibinfo {title} {Nevanlinna analytical continuation},\ }\href
  {https://doi.org/10.1103/PhysRevLett.126.056402} {\bibfield  {journal}
  {\bibinfo  {journal} {Phys. Rev. Lett.}\ }\textbf {\bibinfo {volume} {126}},\
  \bibinfo {pages} {056402} (\bibinfo {year} {2021})}\BibitemShut {NoStop}%
\bibitem [{\citenamefont {Bulla}\ \emph {et~al.}(2008)\citenamefont {Bulla},
  \citenamefont {Costi},\ and\ \citenamefont {Pruschke}}]{BullaPruschke2008}%
  \BibitemOpen
  \bibfield  {author} {\bibinfo {author} {\bibfnamefont {R.}~\bibnamefont
  {Bulla}}, \bibinfo {author} {\bibfnamefont {T.~A.}\ \bibnamefont {Costi}},\
  and\ \bibinfo {author} {\bibfnamefont {T.}~\bibnamefont {Pruschke}},\
  }\bibfield  {title} {\bibinfo {title} {Numerical renormalization group method
  for quantum impurity systems},\ }\href
  {https://doi.org/10.1103/RevModPhys.80.395} {\bibfield  {journal} {\bibinfo
  {journal} {Rev. Mod. Phys.}\ }\textbf {\bibinfo {volume} {80}},\ \bibinfo
  {pages} {395} (\bibinfo {year} {2008})}\BibitemShut {NoStop}%
\bibitem [{\citenamefont {Wolf}\ \emph
  {et~al.}(2014{\natexlab{a}})\citenamefont {Wolf}, \citenamefont {McCulloch},
  \citenamefont {Parcollet},\ and\ \citenamefont
  {Schollw\"ock}}]{WolfSchollwock2014b}%
  \BibitemOpen
  \bibfield  {author} {\bibinfo {author} {\bibfnamefont {F.~A.}\ \bibnamefont
  {Wolf}}, \bibinfo {author} {\bibfnamefont {I.~P.}\ \bibnamefont {McCulloch}},
  \bibinfo {author} {\bibfnamefont {O.}~\bibnamefont {Parcollet}},\ and\
  \bibinfo {author} {\bibfnamefont {U.}~\bibnamefont {Schollw\"ock}},\
  }\bibfield  {title} {\bibinfo {title} {Chebyshev matrix product state
  impurity solver for dynamical mean-field theory},\ }\href
  {https://doi.org/10.1103/PhysRevB.90.115124} {\bibfield  {journal} {\bibinfo
  {journal} {Phys. Rev. B}\ }\textbf {\bibinfo {volume} {90}},\ \bibinfo
  {pages} {115124} (\bibinfo {year} {2014}{\natexlab{a}})}\BibitemShut
  {NoStop}%
\bibitem [{\citenamefont {Wolf}\ \emph
  {et~al.}(2014{\natexlab{b}})\citenamefont {Wolf}, \citenamefont {McCulloch},\
  and\ \citenamefont {Schollw\"ock}}]{WolfSchollwock2014}%
  \BibitemOpen
  \bibfield  {author} {\bibinfo {author} {\bibfnamefont {F.~A.}\ \bibnamefont
  {Wolf}}, \bibinfo {author} {\bibfnamefont {I.~P.}\ \bibnamefont
  {McCulloch}},\ and\ \bibinfo {author} {\bibfnamefont {U.}~\bibnamefont
  {Schollw\"ock}},\ }\bibfield  {title} {\bibinfo {title} {Solving
  nonequilibrium dynamical mean-field theory using matrix product states},\
  }\href {https://doi.org/10.1103/PhysRevB.90.235131} {\bibfield  {journal}
  {\bibinfo  {journal} {Phys. Rev. B}\ }\textbf {\bibinfo {volume} {90}},\
  \bibinfo {pages} {235131} (\bibinfo {year} {2014}{\natexlab{b}})}\BibitemShut
  {NoStop}%
\bibitem [{\citenamefont {Ganahl}\ \emph {et~al.}(2014)\citenamefont {Ganahl},
  \citenamefont {Thunstr\"om}, \citenamefont {Verstraete}, \citenamefont
  {Held},\ and\ \citenamefont {Evertz}}]{GanahlEvertz2014}%
  \BibitemOpen
  \bibfield  {author} {\bibinfo {author} {\bibfnamefont {M.}~\bibnamefont
  {Ganahl}}, \bibinfo {author} {\bibfnamefont {P.}~\bibnamefont {Thunstr\"om}},
  \bibinfo {author} {\bibfnamefont {F.}~\bibnamefont {Verstraete}}, \bibinfo
  {author} {\bibfnamefont {K.}~\bibnamefont {Held}},\ and\ \bibinfo {author}
  {\bibfnamefont {H.~G.}\ \bibnamefont {Evertz}},\ }\bibfield  {title}
  {\bibinfo {title} {Chebyshev expansion for impurity models using matrix
  product states},\ }\href {https://doi.org/10.1103/PhysRevB.90.045144}
  {\bibfield  {journal} {\bibinfo  {journal} {Phys. Rev. B}\ }\textbf {\bibinfo
  {volume} {90}},\ \bibinfo {pages} {045144} (\bibinfo {year}
  {2014})}\BibitemShut {NoStop}%
\bibitem [{\citenamefont {Ganahl}\ \emph {et~al.}(2015)\citenamefont {Ganahl},
  \citenamefont {Aichhorn}, \citenamefont {Evertz}, \citenamefont
  {Thunstr\"om}, \citenamefont {Held},\ and\ \citenamefont
  {Verstraete}}]{GanahlVerstraete2015}%
  \BibitemOpen
  \bibfield  {author} {\bibinfo {author} {\bibfnamefont {M.}~\bibnamefont
  {Ganahl}}, \bibinfo {author} {\bibfnamefont {M.}~\bibnamefont {Aichhorn}},
  \bibinfo {author} {\bibfnamefont {H.~G.}\ \bibnamefont {Evertz}}, \bibinfo
  {author} {\bibfnamefont {P.}~\bibnamefont {Thunstr\"om}}, \bibinfo {author}
  {\bibfnamefont {K.}~\bibnamefont {Held}},\ and\ \bibinfo {author}
  {\bibfnamefont {F.}~\bibnamefont {Verstraete}},\ }\bibfield  {title}
  {\bibinfo {title} {Efficient DMFT impurity solver using real-time dynamics
  with matrix product states},\ }\href
  {https://doi.org/10.1103/PhysRevB.92.155132} {\bibfield  {journal} {\bibinfo
  {journal} {Phys. Rev. B}\ }\textbf {\bibinfo {volume} {92}},\ \bibinfo
  {pages} {155132} (\bibinfo {year} {2015})}\BibitemShut {NoStop}%
\bibitem [{\citenamefont {Kohn}\ and\ \citenamefont
  {Santoro}(2021)}]{KohnSantoro2021}%
  \BibitemOpen
  \bibfield  {author} {\bibinfo {author} {\bibfnamefont {L.}~\bibnamefont
  {Kohn}}\ and\ \bibinfo {author} {\bibfnamefont {G.~E.}\ \bibnamefont
  {Santoro}},\ }\bibfield  {title} {\bibinfo {title} {Efficient mapping for
  anderson impurity problems with matrix product states},\ }\href
  {https://doi.org/10.1103/PhysRevB.104.014303} {\bibfield  {journal} {\bibinfo
   {journal} {Phys. Rev. B}\ }\textbf {\bibinfo {volume} {104}},\ \bibinfo
  {pages} {014303} (\bibinfo {year} {2021})}\BibitemShut {NoStop}%
\bibitem [{\citenamefont {Kohn}\ and\ \citenamefont
  {Santoro}(2022)}]{KohnSantoro2022}%
  \BibitemOpen
  \bibfield  {author} {\bibinfo {author} {\bibfnamefont {L.}~\bibnamefont
  {Kohn}}\ and\ \bibinfo {author} {\bibfnamefont {G.~E.}\ \bibnamefont
  {Santoro}},\ }\bibfield  {title} {\bibinfo {title} {Quench dynamics of the
  anderson impurity model at finite temperature using matrix product states:
  entanglement and bath dynamics},\ }\href
  {https://doi.org/10.1088/1742-5468/ac729b} {\bibfield  {journal} {\bibinfo
  {journal} {J. Stat. Mech. Theory Exp}\ }\textbf {\bibinfo {volume} {2022}},\
  \bibinfo {pages} {063102} (\bibinfo {year} {2022})}\BibitemShut {NoStop}%
\bibitem [{\citenamefont {Mitchell}\ \emph {et~al.}(2014)\citenamefont
  {Mitchell}, \citenamefont {Galpin}, \citenamefont {Wilson-Fletcher},
  \citenamefont {Logan},\ and\ \citenamefont {Bulla}}]{MitchellBulla2014}%
  \BibitemOpen
  \bibfield  {author} {\bibinfo {author} {\bibfnamefont {A.~K.}\ \bibnamefont
  {Mitchell}}, \bibinfo {author} {\bibfnamefont {M.~R.}\ \bibnamefont
  {Galpin}}, \bibinfo {author} {\bibfnamefont {S.}~\bibnamefont
  {Wilson-Fletcher}}, \bibinfo {author} {\bibfnamefont {D.~E.}\ \bibnamefont
  {Logan}},\ and\ \bibinfo {author} {\bibfnamefont {R.}~\bibnamefont {Bulla}},\
  }\bibfield  {title} {\bibinfo {title} {Generalized wilson chain for solving
  multichannel quantum impurity problems},\ }\href
  {https://doi.org/10.1103/PhysRevB.89.121105} {\bibfield  {journal} {\bibinfo
  {journal} {Phys. Rev. B}\ }\textbf {\bibinfo {volume} {89}},\ \bibinfo
  {pages} {121105} (\bibinfo {year} {2014})}\BibitemShut {NoStop}%
\bibitem [{\citenamefont {Stadler}\ \emph {et~al.}(2015)\citenamefont
  {Stadler}, \citenamefont {Yin}, \citenamefont {von Delft}, \citenamefont
  {Kotliar},\ and\ \citenamefont {Weichselbaum}}]{StadlerWeichselbaum2015}%
  \BibitemOpen
  \bibfield  {author} {\bibinfo {author} {\bibfnamefont {K.~M.}\ \bibnamefont
  {Stadler}}, \bibinfo {author} {\bibfnamefont {Z.~P.}\ \bibnamefont {Yin}},
  \bibinfo {author} {\bibfnamefont {J.}~\bibnamefont {von Delft}}, \bibinfo
  {author} {\bibfnamefont {G.}~\bibnamefont {Kotliar}},\ and\ \bibinfo {author}
  {\bibfnamefont {A.}~\bibnamefont {Weichselbaum}},\ }\bibfield  {title}
  {\bibinfo {title} {Dynamical mean-field theory plus numerical
  renormalization-group study of spin-orbital separation in a three-band hund
  metal},\ }\href {https://doi.org/10.1103/PhysRevLett.115.136401} {\bibfield
  {journal} {\bibinfo  {journal} {Phys. Rev. Lett.}\ }\textbf {\bibinfo
  {volume} {115}},\ \bibinfo {pages} {136401} (\bibinfo {year}
  {2015})}\BibitemShut {NoStop}%
\bibitem [{\citenamefont {Horvat}\ \emph {et~al.}(2016)\citenamefont {Horvat},
  \citenamefont {\ifmmode~\check{Z}\else \v{Z}\fi{}itko},\ and\ \citenamefont
  {Mravlje}}]{HorvatMravlje2016}%
  \BibitemOpen
  \bibfield  {author} {\bibinfo {author} {\bibfnamefont {A.}~\bibnamefont
  {Horvat}}, \bibinfo {author} {\bibfnamefont {R.}~\bibnamefont
  {\ifmmode~\check{Z}\else \v{Z}\fi{}itko}},\ and\ \bibinfo {author}
  {\bibfnamefont {J.}~\bibnamefont {Mravlje}},\ }\bibfield  {title} {\bibinfo
  {title} {Low-energy physics of three-orbital impurity model with kanamori
  interaction},\ }\href {https://doi.org/10.1103/PhysRevB.94.165140} {\bibfield
   {journal} {\bibinfo  {journal} {Phys. Rev. B}\ }\textbf {\bibinfo {volume}
  {94}},\ \bibinfo {pages} {165140} (\bibinfo {year} {2016})}\BibitemShut
  {NoStop}%
\bibitem [{\citenamefont {Kugler}\ \emph {et~al.}(2020)\citenamefont {Kugler},
  \citenamefont {Zingl}, \citenamefont {Strand}, \citenamefont {Lee},
  \citenamefont {von Delft},\ and\ \citenamefont
  {Georges}}]{KuglerGeorges2020}%
  \BibitemOpen
  \bibfield  {author} {\bibinfo {author} {\bibfnamefont {F.~B.}\ \bibnamefont
  {Kugler}}, \bibinfo {author} {\bibfnamefont {M.}~\bibnamefont {Zingl}},
  \bibinfo {author} {\bibfnamefont {H.~U.~R.}\ \bibnamefont {Strand}}, \bibinfo
  {author} {\bibfnamefont {S.-S.~B.}\ \bibnamefont {Lee}}, \bibinfo {author}
  {\bibfnamefont {J.}~\bibnamefont {von Delft}},\ and\ \bibinfo {author}
  {\bibfnamefont {A.}~\bibnamefont {Georges}},\ }\bibfield  {title} {\bibinfo
  {title} {Strongly correlated materials from a numerical renormalization group
  perspective: How the fermi-liquid state of
  ${\mathrm{Sr}}_{2}{\mathrm{RuO}}_{4}$ emerges},\ }\href
  {https://doi.org/10.1103/PhysRevLett.124.016401} {\bibfield  {journal}
  {\bibinfo  {journal} {Phys. Rev. Lett.}\ }\textbf {\bibinfo {volume} {124}},\
  \bibinfo {pages} {016401} (\bibinfo {year} {2020})}\BibitemShut {NoStop}%
\bibitem [{\citenamefont {Feynman}\ and\ \citenamefont
  {Vernon}(1963)}]{FeynmanVernon1963}%
  \BibitemOpen
  \bibfield  {author} {\bibinfo {author} {\bibfnamefont {R.~P.}\ \bibnamefont
  {Feynman}}\ and\ \bibinfo {author} {\bibfnamefont {F.~L.}\ \bibnamefont
  {Vernon}},\ }\bibfield  {title} {\bibinfo {title} {The theory of a general
  quantum system interacting with a linear dissipative system},\ }\href
  {https://doi.org/10.1016/0003-4916(63)90068-X} {\bibfield  {journal}
  {\bibinfo  {journal} {Ann. Phys.}\ }\textbf {\bibinfo {volume} {24}},\
  \bibinfo {pages} {118} (\bibinfo {year} {1963})}\BibitemShut {NoStop}%
\bibitem [{\citenamefont {Strathearn}\ \emph {et~al.}(2018)\citenamefont
  {Strathearn}, \citenamefont {Kirton}, \citenamefont {Kilda}, \citenamefont
  {Keeling},\ and\ \citenamefont {Lovett}}]{StrathearnLovett2018}%
  \BibitemOpen
  \bibfield  {author} {\bibinfo {author} {\bibfnamefont {A.}~\bibnamefont
  {Strathearn}}, \bibinfo {author} {\bibfnamefont {P.}~\bibnamefont {Kirton}},
  \bibinfo {author} {\bibfnamefont {D.}~\bibnamefont {Kilda}}, \bibinfo
  {author} {\bibfnamefont {J.}~\bibnamefont {Keeling}},\ and\ \bibinfo {author}
  {\bibfnamefont {B.~W.}\ \bibnamefont {Lovett}},\ }\bibfield  {title}
  {\bibinfo {title} {Efficient non-markovian quantum dynamics using
  time-evolving matrix product operators},\ }\href
  {https://doi.org/10.1038/s41467-018-05617-3} {\bibfield  {journal} {\bibinfo
  {journal} {Nat. Commun.}\ }\textbf {\bibinfo {volume} {9}},\ \bibinfo {pages}
  {3322} (\bibinfo {year} {2018})}\BibitemShut {NoStop}%
\bibitem [{\citenamefont {Schollwöck}(2011)}]{Schollwock2011}%
  \BibitemOpen
  \bibfield  {author} {\bibinfo {author} {\bibfnamefont {U.}~\bibnamefont
  {Schollwöck}},\ }\bibfield  {title} {\bibinfo {title} {The density-matrix
  renormalization group in the age of matrix product states},\ }\href
  {https://doi.org/https://doi.org/10.1016/j.aop.2010.09.012} {\bibfield
  {journal} {\bibinfo  {journal} {Ann. Phys.}\ }\textbf {\bibinfo {volume}
  {326}},\ \bibinfo {pages} {96} (\bibinfo {year} {2011})}\BibitemShut
  {NoStop}%
\bibitem [{\citenamefont {J{\o}rgensen}\ and\ \citenamefont
  {Pollock}(2019)}]{joergensen2019-exploiting}%
  \BibitemOpen
  \bibfield  {author} {\bibinfo {author} {\bibfnamefont {M.~R.}\ \bibnamefont
  {J{\o}rgensen}}\ and\ \bibinfo {author} {\bibfnamefont {F.~A.}\ \bibnamefont
  {Pollock}},\ }\bibfield  {title} {\bibinfo {title} {Exploiting the causal
  tensor network structure of quantum processes to efficiently simulate
  non-markovian path integrals},\ }\href
  {https://doi.org/10.1103/physrevlett.123.240602} {\bibfield  {journal}
  {\bibinfo  {journal} {Phys. Rev. Lett.}\ }\textbf {\bibinfo {volume} {123}},\
  \bibinfo {pages} {240602} (\bibinfo {year} {2019})}\BibitemShut {NoStop}%
\bibitem [{\citenamefont {Popovic}\ \emph {et~al.}(2021)\citenamefont
  {Popovic}, \citenamefont {Mitchison}, \citenamefont {Strathearn},
  \citenamefont {Lovett}, \citenamefont {Goold},\ and\ \citenamefont
  {Eastham}}]{popovic2021-quantum}%
  \BibitemOpen
  \bibfield  {author} {\bibinfo {author} {\bibfnamefont {M.}~\bibnamefont
  {Popovic}}, \bibinfo {author} {\bibfnamefont {M.~T.}\ \bibnamefont
  {Mitchison}}, \bibinfo {author} {\bibfnamefont {A.}~\bibnamefont
  {Strathearn}}, \bibinfo {author} {\bibfnamefont {B.~W.}\ \bibnamefont
  {Lovett}}, \bibinfo {author} {\bibfnamefont {J.}~\bibnamefont {Goold}},\ and\
  \bibinfo {author} {\bibfnamefont {P.~R.}\ \bibnamefont {Eastham}},\
  }\bibfield  {title} {\bibinfo {title} {Quantum heat statistics with
  time-evolving matrix product operators},\ }\href
  {https://doi.org/10.1103/prxquantum.2.020338} {\bibfield  {journal} {\bibinfo
   {journal} {PRX Quantum}\ }\textbf {\bibinfo {volume} {2}},\ \bibinfo {pages}
  {020338} (\bibinfo {year} {2021})}\BibitemShut {NoStop}%
\bibitem [{\citenamefont {Fux}\ \emph {et~al.}(2021)\citenamefont {Fux},
  \citenamefont {Butler}, \citenamefont {Eastham}, \citenamefont {Lovett},\
  and\ \citenamefont {Keeling}}]{fux2021-efficient}%
  \BibitemOpen
  \bibfield  {author} {\bibinfo {author} {\bibfnamefont {G.~E.}\ \bibnamefont
  {Fux}}, \bibinfo {author} {\bibfnamefont {E.~P.}\ \bibnamefont {Butler}},
  \bibinfo {author} {\bibfnamefont {P.~R.}\ \bibnamefont {Eastham}}, \bibinfo
  {author} {\bibfnamefont {B.~W.}\ \bibnamefont {Lovett}},\ and\ \bibinfo
  {author} {\bibfnamefont {J.}~\bibnamefont {Keeling}},\ }\bibfield  {title}
  {\bibinfo {title} {Efficient exploration of hamiltonian parameter space for
  optimal control of non-markovian open quantum systems},\ }\href
  {https://doi.org/10.1103/physrevlett.126.200401} {\bibfield  {journal}
  {\bibinfo  {journal} {Phys. Rev. Lett.}\ }\textbf {\bibinfo {volume} {126}},\
  \bibinfo {pages} {200401} (\bibinfo {year} {2021})}\BibitemShut {NoStop}%
\bibitem [{\citenamefont {Gribben}\ \emph {et~al.}(2021)\citenamefont
  {Gribben}, \citenamefont {Strathearn}, \citenamefont {Fux}, \citenamefont
  {Kirton},\ and\ \citenamefont {Lovett}}]{gribben2021-using}%
  \BibitemOpen
  \bibfield  {author} {\bibinfo {author} {\bibfnamefont {D.}~\bibnamefont
  {Gribben}}, \bibinfo {author} {\bibfnamefont {A.}~\bibnamefont {Strathearn}},
  \bibinfo {author} {\bibfnamefont {G.~E.}\ \bibnamefont {Fux}}, \bibinfo
  {author} {\bibfnamefont {P.}~\bibnamefont {Kirton}},\ and\ \bibinfo {author}
  {\bibfnamefont {B.~W.}\ \bibnamefont {Lovett}},\ }\bibfield  {title}
  {\bibinfo {title} {Using the environment to understand non-markovian open
  quantum systems},\ }\href {https://doi.org/10.22331/q-2022-10-25-847}
  {\bibfield  {journal} {\bibinfo  {journal} {Quantum}\ }\textbf {\bibinfo
  {volume} {6}},\ \bibinfo {pages} {847} (\bibinfo {year} {2021})}\BibitemShut
  {NoStop}%
\bibitem [{\citenamefont {Otterpohl}\ \emph {et~al.}(2022)\citenamefont
  {Otterpohl}, \citenamefont {Nalbach},\ and\ \citenamefont
  {Thorwart}}]{otterpohl2022-hidden}%
  \BibitemOpen
  \bibfield  {author} {\bibinfo {author} {\bibfnamefont {F.}~\bibnamefont
  {Otterpohl}}, \bibinfo {author} {\bibfnamefont {P.}~\bibnamefont {Nalbach}},\
  and\ \bibinfo {author} {\bibfnamefont {M.}~\bibnamefont {Thorwart}},\
  }\bibfield  {title} {\bibinfo {title} {Hidden phase of the spin-boson
  model},\ }\href {https://doi.org/10.1103/physrevlett.129.120406} {\bibfield
  {journal} {\bibinfo  {journal} {Phys. Rev. Lett.}\ }\textbf {\bibinfo
  {volume} {129}},\ \bibinfo {pages} {120406} (\bibinfo {year}
  {2022})}\BibitemShut {NoStop}%
\bibitem [{\citenamefont {Gribben}\ \emph {et~al.}(2022)\citenamefont
  {Gribben}, \citenamefont {Rouse}, \citenamefont {Iles-Smith}, \citenamefont
  {Strathearn}, \citenamefont {Maguire}, \citenamefont {Kirton}, \citenamefont
  {Nazir}, \citenamefont {Gauger},\ and\ \citenamefont
  {Lovett}}]{gribben2022-exact}%
  \BibitemOpen
  \bibfield  {author} {\bibinfo {author} {\bibfnamefont {D.}~\bibnamefont
  {Gribben}}, \bibinfo {author} {\bibfnamefont {D.~M.}\ \bibnamefont {Rouse}},
  \bibinfo {author} {\bibfnamefont {J.}~\bibnamefont {Iles-Smith}}, \bibinfo
  {author} {\bibfnamefont {A.}~\bibnamefont {Strathearn}}, \bibinfo {author}
  {\bibfnamefont {H.}~\bibnamefont {Maguire}}, \bibinfo {author} {\bibfnamefont
  {P.}~\bibnamefont {Kirton}}, \bibinfo {author} {\bibfnamefont
  {A.}~\bibnamefont {Nazir}}, \bibinfo {author} {\bibfnamefont {E.~M.}\
  \bibnamefont {Gauger}},\ and\ \bibinfo {author} {\bibfnamefont {B.~W.}\
  \bibnamefont {Lovett}},\ }\bibfield  {title} {\bibinfo {title} {Exact
  dynamics of nonadditive environments in non-markovian open quantum systems},\
  }\href {https://doi.org/10.1103/prxquantum.3.010321} {\bibfield  {journal}
  {\bibinfo  {journal} {PRX Quantum}\ }\textbf {\bibinfo {volume} {3}},\
  \bibinfo {pages} {010321} (\bibinfo {year} {2022})}\BibitemShut {NoStop}%
\bibitem [{\citenamefont {Thoenniss}\ \emph
  {et~al.}(2023{\natexlab{a}})\citenamefont {Thoenniss}, \citenamefont
  {Lerose},\ and\ \citenamefont {Abanin}}]{ThoennissAbanin2023a}%
  \BibitemOpen
  \bibfield  {author} {\bibinfo {author} {\bibfnamefont {J.}~\bibnamefont
  {Thoenniss}}, \bibinfo {author} {\bibfnamefont {A.}~\bibnamefont {Lerose}},\
  and\ \bibinfo {author} {\bibfnamefont {D.~A.}\ \bibnamefont {Abanin}},\
  }\bibfield  {title} {\bibinfo {title} {Nonequilibrium quantum impurity
  problems via matrix-product states in the temporal domain},\ }\href
  {https://doi.org/10.1103/PhysRevB.107.195101} {\bibfield  {journal} {\bibinfo
   {journal} {Phys. Rev. B}\ }\textbf {\bibinfo {volume} {107}},\ \bibinfo
  {pages} {195101} (\bibinfo {year} {2023}{\natexlab{a}})}\BibitemShut
  {NoStop}%
\bibitem [{\citenamefont {Thoenniss}\ \emph
  {et~al.}(2023{\natexlab{b}})\citenamefont {Thoenniss}, \citenamefont
  {Sonner}, \citenamefont {Lerose},\ and\ \citenamefont
  {Abanin}}]{ThoennissAbanin2023b}%
  \BibitemOpen
  \bibfield  {author} {\bibinfo {author} {\bibfnamefont {J.}~\bibnamefont
  {Thoenniss}}, \bibinfo {author} {\bibfnamefont {M.}~\bibnamefont {Sonner}},
  \bibinfo {author} {\bibfnamefont {A.}~\bibnamefont {Lerose}},\ and\ \bibinfo
  {author} {\bibfnamefont {D.~A.}\ \bibnamefont {Abanin}},\ }\bibfield  {title}
  {\bibinfo {title} {Efficient method for quantum impurity problems out of
  equilibrium},\ }\href {https://doi.org/10.1103/PhysRevB.107.L201115}
  {\bibfield  {journal} {\bibinfo  {journal} {Phys. Rev. B}\ }\textbf {\bibinfo
  {volume} {107}},\ \bibinfo {pages} {L201115} (\bibinfo {year}
  {2023}{\natexlab{b}})}\BibitemShut {NoStop}%
\bibitem [{\citenamefont {Fishman}\ and\ \citenamefont
  {White}(2015)}]{fishman2015-compression}%
  \BibitemOpen
  \bibfield  {author} {\bibinfo {author} {\bibfnamefont {M.~T.}\ \bibnamefont
  {Fishman}}\ and\ \bibinfo {author} {\bibfnamefont {S.~R.}\ \bibnamefont
  {White}},\ }\bibfield  {title} {\bibinfo {title} {Compression of correlation
  matrices and an efficient method for forming matrix product states of
  fermionic gaussian states},\ }\href
  {https://doi.org/10.1103/physrevb.92.075132} {\bibfield  {journal} {\bibinfo
  {journal} {Phys. Rev. B}\ }\textbf {\bibinfo {volume} {92}},\ \bibinfo
  {pages} {075132} (\bibinfo {year} {2015})}\BibitemShut {NoStop}%
\bibitem [{\citenamefont {Kamenev}\ and\ \citenamefont
  {Levchenko}(2009)}]{kamenev2009-keldysh}%
  \BibitemOpen
  \bibfield  {author} {\bibinfo {author} {\bibfnamefont {A.}~\bibnamefont
  {Kamenev}}\ and\ \bibinfo {author} {\bibfnamefont {A.}~\bibnamefont
  {Levchenko}},\ }\bibfield  {title} {\bibinfo {title} {Keldysh technique and
  non-linear $\sigma$-model: Basic principles and applications},\ }\href
  {https://doi.org/10.1080/00018730902850504} {\bibfield  {journal} {\bibinfo
  {journal} {Adv. Phys.}\ }\textbf {\bibinfo {volume} {58}},\ \bibinfo {pages}
  {197} (\bibinfo {year} {2009})}\BibitemShut {NoStop}%
\bibitem [{\citenamefont {Negele}\ and\ \citenamefont
  {Orland}(1998)}]{negele1998-quantum}%
  \BibitemOpen
  \bibfield  {author} {\bibinfo {author} {\bibfnamefont {J.~W.}\ \bibnamefont
  {Negele}}\ and\ \bibinfo {author} {\bibfnamefont {H.}~\bibnamefont
  {Orland}},\ }\href@noop {} {\emph {\bibinfo {title} {Quantum Many-Particle
  Systems}}}\ (\bibinfo  {publisher} {Westview Press},\ \bibinfo {year}
  {1998})\BibitemShut {NoStop}%
\bibitem [{\citenamefont {Trotter}(1959)}]{trotter1959-product}%
  \BibitemOpen
  \bibfield  {author} {\bibinfo {author} {\bibfnamefont {H.~F.}\ \bibnamefont
  {Trotter}},\ }\bibfield  {title} {\bibinfo {title} {On the product of
  semi-groups of operators},\ }\href
  {https://doi.org/10.1090/s0002-9939-1959-0108732-6} {\bibfield  {journal}
  {\bibinfo  {journal} {Proc. Am. Math. Soc.}\ }\textbf {\bibinfo {volume}
  {10}},\ \bibinfo {pages} {545} (\bibinfo {year} {1959})}\BibitemShut
  {NoStop}%
\bibitem [{\citenamefont {Suzuki}(1976)}]{suzuki1976-generalized}%
  \BibitemOpen
  \bibfield  {author} {\bibinfo {author} {\bibfnamefont {M.}~\bibnamefont
  {Suzuki}},\ }\bibfield  {title} {\bibinfo {title} {Generalized trotter's
  formula and systematic approximants of exponential operators and inner
  derivations with applications to many-body problems},\ }\href
  {https://doi.org/10.1007/bf01609348} {\bibfield  {journal} {\bibinfo
  {journal} {Commun. Math. Phys.}\ }\textbf {\bibinfo {volume} {51}},\ \bibinfo
  {pages} {183} (\bibinfo {year} {1976})}\BibitemShut {NoStop}%
\bibitem [{\citenamefont {Makarov}\ and\ \citenamefont
  {Makri}(1994)}]{makarov1994-path}%
  \BibitemOpen
  \bibfield  {author} {\bibinfo {author} {\bibfnamefont {D.~E.}\ \bibnamefont
  {Makarov}}\ and\ \bibinfo {author} {\bibfnamefont {N.}~\bibnamefont
  {Makri}},\ }\bibfield  {title} {\bibinfo {title} {Path integrals for
  dissipative systems by tensor multiplication. condensed phase quantum
  dynamics for arbitrarily long time},\ }\href
  {https://doi.org/10.1016/0009-2614(94)00275-4} {\bibfield  {journal}
  {\bibinfo  {journal} {Chem. Phys. Lett.}\ }\textbf {\bibinfo {volume}
  {221}},\ \bibinfo {pages} {482} (\bibinfo {year} {1994})}\BibitemShut
  {NoStop}%
\bibitem [{\citenamefont {Makri}(1995)}]{makri1995-numerical}%
  \BibitemOpen
  \bibfield  {author} {\bibinfo {author} {\bibfnamefont {N.}~\bibnamefont
  {Makri}},\ }\bibfield  {title} {\bibinfo {title} {Numerical path integral
  techniques for long time dynamics of quantum dissipative systems},\ }\href
  {https://doi.org/10.1063/1.531046} {\bibfield  {journal} {\bibinfo  {journal}
  {J. Math. Phys.}\ }\textbf {\bibinfo {volume} {36}},\ \bibinfo {pages} {2430}
  (\bibinfo {year} {1995})}\BibitemShut {NoStop}%
\bibitem [{\citenamefont {Dattani}\ \emph {et~al.}(2012)\citenamefont
  {Dattani}, \citenamefont {Pollock},\ and\ \citenamefont
  {Wilkins}}]{dattani2012-analytic}%
  \BibitemOpen
  \bibfield  {author} {\bibinfo {author} {\bibfnamefont {N.~S.}\ \bibnamefont
  {Dattani}}, \bibinfo {author} {\bibfnamefont {F.~A.}\ \bibnamefont
  {Pollock}},\ and\ \bibinfo {author} {\bibfnamefont {D.~M.}\ \bibnamefont
  {Wilkins}},\ }\bibfield  {title} {\bibinfo {title} {Analytic influence
  functionals for numerical feynman integrals in most open quantum systems},\
  }\href {http://www.naturalspublishing.com/Article.asp?ArtcID=407} {\bibfield
  {journal} {\bibinfo  {journal} {Quantum Phys. Lett.}\ }\textbf {\bibinfo
  {volume} {1}},\ \bibinfo {pages} {35} (\bibinfo {year} {2012})}\BibitemShut
  {NoStop}%
\bibitem [{\citenamefont {Ng}\ \emph {et~al.}(2023)\citenamefont {Ng},
  \citenamefont {Park}, \citenamefont {Millis}, \citenamefont {Chan},\ and\
  \citenamefont {Reichman}}]{NgReichman2023}%
  \BibitemOpen
  \bibfield  {author} {\bibinfo {author} {\bibfnamefont {N.}~\bibnamefont
  {Ng}}, \bibinfo {author} {\bibfnamefont {G.}~\bibnamefont {Park}}, \bibinfo
  {author} {\bibfnamefont {A.~J.}\ \bibnamefont {Millis}}, \bibinfo {author}
  {\bibfnamefont {G.~K.-L.}\ \bibnamefont {Chan}},\ and\ \bibinfo {author}
  {\bibfnamefont {D.~R.}\ \bibnamefont {Reichman}},\ }\bibfield  {title}
  {\bibinfo {title} {Real-time evolution of anderson impurity models via tensor
  network influence functionals},\ }\href
  {https://doi.org/10.1103/PhysRevB.107.125103} {\bibfield  {journal} {\bibinfo
   {journal} {Phys. Rev. B}\ }\textbf {\bibinfo {volume} {107}},\ \bibinfo
  {pages} {125103} (\bibinfo {year} {2023})}\BibitemShut {NoStop}%
\bibitem [{\citenamefont {Yosprakob}(2023)}]{Yosprakob2023}%
  \BibitemOpen
  \bibfield  {author} {\bibinfo {author} {\bibfnamefont {A.}~\bibnamefont
  {Yosprakob}},\ }\bibfield  {title} {\bibinfo {title} {{GrassmannTN: A Python
  package for Grassmann tensor network computations}},\ }\href
  {https://doi.org/10.21468/SciPostPhysCodeb.20} {\bibfield  {journal}
  {\bibinfo  {journal} {SciPost Phys. Codebases}\ ,\ \bibinfo {pages} {20}}
  (\bibinfo {year} {2023})}\BibitemShut {NoStop}%
\bibitem [{\citenamefont {Singh}\ \emph {et~al.}(2011)\citenamefont {Singh},
  \citenamefont {Pfeifer},\ and\ \citenamefont {Vidal}}]{SinghVidal2011}%
  \BibitemOpen
  \bibfield  {author} {\bibinfo {author} {\bibfnamefont {S.}~\bibnamefont
  {Singh}}, \bibinfo {author} {\bibfnamefont {R.~N.~C.}\ \bibnamefont
  {Pfeifer}},\ and\ \bibinfo {author} {\bibfnamefont {G.}~\bibnamefont
  {Vidal}},\ }\bibfield  {title} {\bibinfo {title} {Tensor network states and
  algorithms in the presence of a global U(1) symmetry},\ }\href
  {https://doi.org/10.1103/PhysRevB.83.115125} {\bibfield  {journal} {\bibinfo
  {journal} {Phys. Rev. B}\ }\textbf {\bibinfo {volume} {83}},\ \bibinfo
  {pages} {115125} (\bibinfo {year} {2011})}\BibitemShut {NoStop}%
\bibitem [{\citenamefont {Fidkowski}\ and\ \citenamefont
  {Kitaev}(2011)}]{FidkowskiKitaev2011}%
  \BibitemOpen
  \bibfield  {author} {\bibinfo {author} {\bibfnamefont {L.}~\bibnamefont
  {Fidkowski}}\ and\ \bibinfo {author} {\bibfnamefont {A.}~\bibnamefont
  {Kitaev}},\ }\bibfield  {title} {\bibinfo {title} {Topological phases of
  fermions in one dimension},\ }\href
  {https://doi.org/10.1103/PhysRevB.83.075103} {\bibfield  {journal} {\bibinfo
  {journal} {Phys. Rev. B}\ }\textbf {\bibinfo {volume} {83}},\ \bibinfo
  {pages} {075103} (\bibinfo {year} {2011})}\BibitemShut {NoStop}%
\bibitem [{\citenamefont {Bultinck}\ \emph {et~al.}(2017)\citenamefont
  {Bultinck}, \citenamefont {Williamson}, \citenamefont {Haegeman},\ and\
  \citenamefont {Verstraete}}]{bultinck2017-fermionic}%
  \BibitemOpen
  \bibfield  {author} {\bibinfo {author} {\bibfnamefont {N.}~\bibnamefont
  {Bultinck}}, \bibinfo {author} {\bibfnamefont {D.~J.}\ \bibnamefont
  {Williamson}}, \bibinfo {author} {\bibfnamefont {J.}~\bibnamefont
  {Haegeman}},\ and\ \bibinfo {author} {\bibfnamefont {F.}~\bibnamefont
  {Verstraete}},\ }\bibfield  {title} {\bibinfo {title} {Fermionic matrix
  product states and one-dimensional topological phases},\ }\href
  {https://doi.org/10.1103/PhysRevB.95.075108} {\bibfield  {journal} {\bibinfo
  {journal} {Phys. Rev. B}\ }\textbf {\bibinfo {volume} {95}},\ \bibinfo
  {pages} {075108} (\bibinfo {year} {2017})}\BibitemShut {NoStop}%
\bibitem [{\citenamefont {Gu}(2013)}]{gu2013-efficient}%
  \BibitemOpen
  \bibfield  {author} {\bibinfo {author} {\bibfnamefont {Z.-C.}\ \bibnamefont
  {Gu}},\ }\bibfield  {title} {\bibinfo {title} {Efficient simulation of
  grassmann tensor product states},\ }\href
  {https://doi.org/10.1103/physrevb.88.115139} {\bibfield  {journal} {\bibinfo
  {journal} {Phys. Rev. B}\ }\textbf {\bibinfo {volume} {88}},\ \bibinfo
  {pages} {115139} (\bibinfo {year} {2013})}\BibitemShut {NoStop}%
\bibitem [{\citenamefont {Yoshimura}\ \emph {et~al.}(2018)\citenamefont
  {Yoshimura}, \citenamefont {Kuramashi}, \citenamefont {Nakamura},
  \citenamefont {Takeda},\ and\ \citenamefont
  {Sakai}}]{yoshimura2018-calculation}%
  \BibitemOpen
  \bibfield  {author} {\bibinfo {author} {\bibfnamefont {Y.}~\bibnamefont
  {Yoshimura}}, \bibinfo {author} {\bibfnamefont {Y.}~\bibnamefont
  {Kuramashi}}, \bibinfo {author} {\bibfnamefont {Y.}~\bibnamefont {Nakamura}},
  \bibinfo {author} {\bibfnamefont {S.}~\bibnamefont {Takeda}},\ and\ \bibinfo
  {author} {\bibfnamefont {R.}~\bibnamefont {Sakai}},\ }\bibfield  {title}
  {\bibinfo {title} {Calculation of fermionic green functions with grassmann
  higher-order tensor renormalization group},\ }\href
  {https://doi.org/10.1103/physrevd.97.054511} {\bibfield  {journal} {\bibinfo
  {journal} {Phys. Rev. D}\ }\textbf {\bibinfo {volume} {97}},\ \bibinfo
  {pages} {054511} (\bibinfo {year} {2018})}\BibitemShut {NoStop}%
\bibitem [{\citenamefont {Akiyama}\ and\ \citenamefont
  {Kadoh}(2021)}]{akiyama2021-more}%
  \BibitemOpen
  \bibfield  {author} {\bibinfo {author} {\bibfnamefont {S.}~\bibnamefont
  {Akiyama}}\ and\ \bibinfo {author} {\bibfnamefont {D.}~\bibnamefont
  {Kadoh}},\ }\bibfield  {title} {\bibinfo {title} {More about the grassmann
  tensor renormalization group},\ }\href
  {https://doi.org/10.1007/JHEP10(2021)188} {\bibfield  {journal} {\bibinfo
  {journal} {J. High Energy Phys.}\ }\textbf {\bibinfo {volume} {2021}},\
  \bibinfo {pages} {188}}\BibitemShut {NoStop}%
\bibitem [{\citenamefont {Ba\~nuls}\ \emph {et~al.}(2009)\citenamefont
  {Ba\~nuls}, \citenamefont {Hastings}, \citenamefont {Verstraete},\ and\
  \citenamefont {Cirac}}]{BanulsCirac2009}%
  \BibitemOpen
  \bibfield  {author} {\bibinfo {author} {\bibfnamefont {M.~C.}\ \bibnamefont
  {Ba\~nuls}}, \bibinfo {author} {\bibfnamefont {M.~B.}\ \bibnamefont
  {Hastings}}, \bibinfo {author} {\bibfnamefont {F.}~\bibnamefont
  {Verstraete}},\ and\ \bibinfo {author} {\bibfnamefont {J.~I.}\ \bibnamefont
  {Cirac}},\ }\bibfield  {title} {\bibinfo {title} {Matrix product states for
  dynamical simulation of infinite chains},\ }\href
  {https://doi.org/10.1103/PhysRevLett.102.240603} {\bibfield  {journal}
  {\bibinfo  {journal} {Phys. Rev. Lett.}\ }\textbf {\bibinfo {volume} {102}},\
  \bibinfo {pages} {240603} (\bibinfo {year} {2009})}\BibitemShut {NoStop}%
\bibitem [{\citenamefont {Paeckel}\ \emph {et~al.}(2019)\citenamefont
  {Paeckel}, \citenamefont {Köhler}, \citenamefont {Swoboda}, \citenamefont
  {Manmana}, \citenamefont {Schollwöck},\ and\ \citenamefont
  {Hubig}}]{PaeckelHubig2019}%
  \BibitemOpen
  \bibfield  {author} {\bibinfo {author} {\bibfnamefont {S.}~\bibnamefont
  {Paeckel}}, \bibinfo {author} {\bibfnamefont {T.}~\bibnamefont {Köhler}},
  \bibinfo {author} {\bibfnamefont {A.}~\bibnamefont {Swoboda}}, \bibinfo
  {author} {\bibfnamefont {S.~R.}\ \bibnamefont {Manmana}}, \bibinfo {author}
  {\bibfnamefont {U.}~\bibnamefont {Schollwöck}},\ and\ \bibinfo {author}
  {\bibfnamefont {C.}~\bibnamefont {Hubig}},\ }\bibfield  {title} {\bibinfo
  {title} {Time-evolution methods for matrix-product states},\ }\href
  {https://doi.org/https://doi.org/10.1016/j.aop.2019.167998} {\bibfield
  {journal} {\bibinfo  {journal} {Ann. Phys.}\ }\textbf {\bibinfo {volume}
  {411}},\ \bibinfo {pages} {167998} (\bibinfo {year} {2019})}\BibitemShut
  {NoStop}%
\bibitem [{\citenamefont {Bertrand}\ \emph {et~al.}(2019)\citenamefont
  {Bertrand}, \citenamefont {Florens}, \citenamefont {Parcollet},\ and\
  \citenamefont {Waintal}}]{BertrandWaintal2019}%
  \BibitemOpen
  \bibfield  {author} {\bibinfo {author} {\bibfnamefont {C.}~\bibnamefont
  {Bertrand}}, \bibinfo {author} {\bibfnamefont {S.}~\bibnamefont {Florens}},
  \bibinfo {author} {\bibfnamefont {O.}~\bibnamefont {Parcollet}},\ and\
  \bibinfo {author} {\bibfnamefont {X.}~\bibnamefont {Waintal}},\ }\bibfield
  {title} {\bibinfo {title} {Reconstructing nonequilibrium regimes of quantum
  many-body systems from the analytical structure of perturbative expansions},\
  }\href {https://doi.org/10.1103/PhysRevX.9.041008} {\bibfield  {journal}
  {\bibinfo  {journal} {Phys. Rev. X}\ }\textbf {\bibinfo {volume} {9}},\
  \bibinfo {pages} {041008} (\bibinfo {year} {2019})}\BibitemShut {NoStop}%
\bibitem [{\citenamefont {Chen}(2023)}]{chen2023-heat}%
  \BibitemOpen
  \bibfield  {author} {\bibinfo {author} {\bibfnamefont {R.}~\bibnamefont
  {Chen}},\ }\bibfield  {title} {\bibinfo {title} {Heat current in
  non-markovian open systems},\ }\href
  {https://doi.org/10.1088/1367-2630/acc60a} {\bibfield  {journal} {\bibinfo
  {journal} {New J. Phys.}\ }\textbf {\bibinfo {volume} {25}},\ \bibinfo
  {pages} {033035} (\bibinfo {year} {2023})}\BibitemShut {NoStop}%
\bibitem [{\citenamefont {Strathearn}(2020)}]{strathearn2020-modelling}%
  \BibitemOpen
  \bibfield  {author} {\bibinfo {author} {\bibfnamefont {A.}~\bibnamefont
  {Strathearn}},\ }\href {https://doi.org/10.1007/978-3-030-54975-6} {\emph
  {\bibinfo {title} {Modelling Non-Markovian Quantum Systems Using Tensor
  Networks}}}\ (\bibinfo  {publisher} {Springer International Publishing, Cham,
  Switzerland},\ \bibinfo {year} {2020})\BibitemShut {NoStop}%
\end{thebibliography}
%

\end{document}